\documentclass[12pt]{article}
\input epsf.sty
\usepackage{mathrsfs}
\usepackage{amsmath}
\usepackage{mathrsfs}
\usepackage{amssymb}
\usepackage{color}
\usepackage{psfrag}

\newcommand{\bea}{\begin{eqnarray}}
\newcommand{\eea}{\end{eqnarray}}
\newcommand{\be}{\begin{equation}}
\newcommand{\ee}{\end{equation}}
\newcommand{\vs}[1]{\vspace{#1 mm}}

\renewcommand{\a}{\alpha}
\renewcommand{\b}{\beta}

\newcommand{\dsl}{\pa \kern-0.5em /}

\newcommand{\pa}{\partial}

\newcommand{\nn}{\nonumber\\}

\newcommand{\eqn}[1]{(\ref{#1})}

\begin{document}
\topmargin 0pt
\oddsidemargin 0mm

\begin{flushright}

USTC-ICTS-10-11\\




\end{flushright}

\vspace{2mm}

\begin{center}

{\Large \bf Phase transitions and critical behavior of
black branes in canonical ensemble}

\vs{10}

{\large J. X. Lu$^a$\footnote{E-mail: jxlu@ustc.edu.cn}, Shibaji
Roy$^b$\footnote{E-mail: shibaji.roy@saha.ac.in} and Zhiguang
Xiao$^a$\footnote{E-mail: xiaozg@ustc.edu.cn}}

 \vspace{4mm}

{\em

 $^a$ Interdisciplinary Center for Theoretical Study\\
 University of Science and Technology of China, Hefei, Anhui
 230026, China\\

\vs{4}

 $^b$ Saha Institute of Nuclear Physics,
 1/AF Bidhannagar, Calcutta-700 064, India\\}

\end{center}

\vs{10}

\begin{abstract}

We study the thermodynamics and phase structure of asymptotically
flat non-dilatonic as well as dilatonic black branes in a cavity in
arbitrary dimensions ($D$). We consider the canonical ensemble and
so the charge inside the cavity and the temperature at the wall are
fixed. We analyze the stability of the black brane equilibrium
states and derive the phase structures. For the zero charge case we
find an analog of Hawking-Page phase transition for these black
branes in arbitrary dimensions. When the charge is non-zero, we find
that below a critical value of the charge, the phase diagram has a
line of first-order phase transition in a certain range of
temperatures which ends up at a second order phase transition point
(critical point) as the charge attains the critical value. We
calculate the critical exponents at that critical point. Although
our discussion is mainly concerned with the non-dilatonic branes, we
show how it easily carries over to the dilatonic branes as well.

\end{abstract}

\newpage

\section{Introduction}

Over the years much is known about the thermodynamics and phase structure
of black holes in asymptotically AdS space. The reason is that the black holes
in AdS space, unlike those in flat space, are thermodynamically
stable \cite{Hawking:1982dh}. Further, more interests were drawn
with the advent of AdS/CFT
correspondence \cite{Maldacena:1997re,Witten:1998qj,Gubser:1998bc} as the
AdS black holes provide a good laboratory for testing
the correspondence at finite temperature \cite{Witten:1998zw}. So, for
example, the AdS black holes
are well-known to undergo a Hawking-Page phase transition
\cite{Hawking:1982dh} and by AdS/CFT this
corresponds to the confinement-deconfinement phase transition in large $N$
gauge theory \cite{Witten:1998zw}. Similarly, the phase structure of
the charged AdS black holes
which includes, in the canonical ensemble, a similarity with the van der
Waals-Maxwell liquid-gas system, can also be understood from the dual field
theories \cite{Chamblin:1999tk,Chamblin:1999hg}.

However, it was pointed out \cite{Carlip:2003ne,Lundgren:2006kt} that
the above mentioned phase structure is not
unique for the asymptotically AdS black holes. In fact, very similar phase
structure was shown to arise for the suitably stabilized asymptotically flat
as well as asymptotically dS black holes \cite{Carlip:2003ne}. As, the higher
dimensional theories
like string or M-theory admits higher dimensional black objects like black
$p$-branes \cite{Horowitz:1991cd,Duff:1993ye,Duff:1996hp}, it is natural
to ask what kind of phase structures do they give
rise to -- do they have a similar phase structure as the black holes or
they have a different phase structure altogether?

Motivated by this we looked into the thermodynamics and the phase
structure of black $p$-brane solution of $D$-dimensional gravity
coupled to $(p+1)$-form gauge field. In the beginning, we consider
only the non-dilatonic branes, (when $D=11$, they correspond to M2
and M5 branes of M-theory and when $D=10$, it is the D3-brane of
string theory) and then towards the end we show how the analysis
carries over to the dilatonic branes as well. The solutions we
consider are as usual asymptotically flat and so they are
thermodynamically unstable and an isolated black brane would radiate
energy in the form of Hawking radiation. In order to restore
thermodynamic stability so that equilibrium thermodynamics and the
phase structure can be studied, we must consider ensembles that
include not only the branes under consideration but also their
environment. As self-gravitating systems are spatially
inhomogeneous, any specification of such ensembles requires not just
thermodynamic quantities of interest but also the place at which
they take the specified values. In other words, we place the brane
in a cavity {\it a la} York \cite{York:1986it} (see also \cite{Brown:1994sn,
Parentani:1994wr,Peca:1998dv,Gregory:2001bd,Zaslavskii:2003is}) 
and its extension in
the charged case \cite{Braden:1990hw}. Concretely, we will keep the
temperature fixed at the surface of the cavity and as the black
$p$-branes are charged under the $(p+1)$-form gauge field, we will
keep the charge inside the cavity also fixed. This will define a
canonical ensemble and we will study the phase structure of the
black $p$-branes in this ensemble.

After some generalities we first consider the case when the charge
in the cavity is fixed to be zero. In this case we find that there
is a minimum temperature below which no black brane state exists in
equilibrium inside the cavity. But above this temperature there
exist two black brane states with different radii. The larger one is
locally stable and the smaller one is unstable. The locally stable
or `supercooled' large black brane will eventually decay to
energetically more favorable state `hot flat space'. There is a
phase transition temperature at which the `hot flat space' and the
corresponding large black brane can coexist\footnote{One expects
this to be a topological first order phase transition since this
involves a topology change as well as an entropy change.}. But when
the temperature rises above this transition temperature larger black
brane becomes globally stable and the `hot flat space' can decay to
the black brane. The small black brane still remains unstable. As
the temperature rises more, the size of the small black brane
decreases and that of the large black brane increases and at
infinite temperature the size of the small black brane goes to
zero\footnote{When the temperature is of the order of Planck scale,
classical gravity description will break down and one must consider
quantum gravity.}
 whereas,
the size of the large black brane coincides with the size of the
cavity. This situation is analogous to the Hawking-Page transition
of AdS black holes or asymptotically flat or dS black holes in a
cavity \cite{Hawking:1982dh,Carlip:2003ne}.

When the charge is non-zero but fixed inside the cavity, the phase
structure becomes more complicated. Here we find that when the
charge $|q|$ is greater than a critical value $|q_c|$, there exists
a globally stable black brane solution at every temperature in
between zero and infinity. However, when $|q| < |q_c|$, there is a
range of temperature, where there exist three black brane solutions
with different radii. The black branes of the smallest and the
largest sizes are locally stable as they correspond to the local
minima of the free energy. On the other hand, the intermediate size
black brane is unstable, corresponding to the maximum of the free
energy, and never exists. The free energies of the largest and the
smallest black branes are not the same and one is greater than the
other which depends on the temperature of the system. However, there
exists a transition temperature, for a given charge $|q|<|q_c|$,
where the two free energies become the same. This is the temperature
where two black brane states the smallest and the largest can
coexist and can make a transition freely from one phase to the other
just like the van der Waals-Maxwell liquid-gas phase transition.
Since this transition involves an entropy change, therefore, it is a
first order phase transition. This structure was also noticed in AdS
\cite{Chamblin:1999tk,Chamblin:1999hg}, asymptotically flat and dS
black holes \cite{Lundgren:2006kt,Carlip:2003ne} in canonical
ensemble. When the temperature, in the range we mentioned, is less
(more) than the transition temperature, the smallest black brane has
lower (higher) free energy than the largest black brane. So, in one
case the smallest black brane is globally stable and in other the
largest black brane is globally stable. There is a first order phase
transition line which depends on the charge $|q|<|q_c|$ and the line
gets shrunk as we increase the charge ending up at a second order
phase transition point (critical point) when $|q|=|q_c|$. We
calculate the critical exponents at this critical point and found
that they have very similar structure as in the black hole cases
with a universal critical exponent \cite{Carlip:2003ne,
Chamblin:1999tk} for the specific heat as $-2/3$. Since the
stability analysis for the black branes, in general, is quite
complicated and extracting exact values of the parameters may not
always be easy, we will illustrate the behavior by numerical
calculations except for some special cases specified later.

This paper is organized as follows. In section 2, we give the
non-dilatonic black $p$-brane solution in space-time dimensions $D$
and derive the action from which we study the stability and analyze
the phase structure. The details of the general stability analysis
is discussed in section 3. The phase structure for the case of zero
charge in the cavity is considered in section 4. Section 5 discusses
the case of non-zero charge. The critical exponents are given in
section 6 and we conclude in section 7. The dilatonic brane cases
are also mentioned and discussed in appropriate places in the
respective section and how the whole analysis of the paper carries
over to the dilatonic branes is discussed in the Appendix.

\section{Black $p$-brane solution and the action}

The black $p$-brane solution was originally constructed as a
solution to the ten dimensional supergravity containing a metric, a
dilaton and a $(p+1)$-form gauge field \cite{Horowitz:1991cd}. This
was generalized to arbitrary dimensions in \cite{Duff:1993ye}. These
solutions are given in Lorentzian signature, but for the purpose of
studying the thermodynamics \cite{York:1986it} we will write the
black $p$-brane solution (without the dilaton field as we will be
studying the non-dilatonic branes, the dilatonic branes will be
considered in the Appendix) in Euclidean signature as (see for
example \cite{Duff:1994an}), \bea\label{blackpbrane} ds^2 &=&
\Delta_+\Delta_-^{-\frac{d}{D-2}} dt^2 + \Delta_-^{\frac{\tilde
d}{D-2}} \sum_{i=1}^{d-1} (dx^i)^2 + \left(\Delta_+
\Delta_-\right)^{-1} d\rho^2 + \rho^2 d\Omega_{{\tilde d}+1}^2\nn
A_{[p+1]} &=& -i \left[\left(\frac{r_-}{r_+}\right)^{\tilde d/2} -
\left(\frac{r_- r_+}{\rho^2}\right)^{\tilde d/2}\right]dt\wedge
dx^1\cdots \wedge dx^p\nn F_{[p+2]} &\equiv &  dA_{[p+1]} = -i
{\tilde d} \frac{\left(r_- r_+\right)^{\tilde d/2}}{\rho^{\tilde d
+1}} d\rho \wedge dt \wedge dx^1 \cdots \wedge dx^p\eea In the above
we have defined, \be\label{Delta} \Delta_{\pm} = 1 -
\left(\frac{r_{\pm}}{\rho}\right)^{\tilde d} \ee where $r_{\pm}$ are
the two parameters characterizing the solution which are related to
the mass and the charge of the black brane. The metric in
\eqn{blackpbrane} has an isometry S$^1$ $\times$ SO($d-1$) $\times$
SO($\tilde d + 2$) and therefore represents a $(d-1) \equiv p$-brane
in Euclidean signature. The total space-time dimension is $D = d +
\tilde d + 2$, where the space transverse to the $p$-brane has the
dimensionality $\tilde d + 2$. A $p$-brane couples to the
$(p+1)$-form gauge field whose form and its field-strength are given
in \eqn{blackpbrane}. It is clear from the Lorentzian form of the
above metric that when $r_- = 0$, it reduces to $D$-dimesional
Schwarzschild solution which has an event horizon at $\rho = r_+$,
whereas, at $\rho = r_-$, there is a curvature singularity. So, the
metric in \eqn{blackpbrane} represents a black $p$-brane only for
$r_+ > r_-$, with $r_+ = r_-$ being its extremal limit. Note that in
the above we have defined the gauge potential with a constant shift,
following \cite{Braden:1990hw}, in such a way that it vanishes on
the horizon so that it is well-defined on the local inertial frame.
For the metric in (\ref{blackpbrane}) to be well defined without a
conical singularity at $\rho = r_+$, the Euclidean time in the
metric is periodic with a periodicity \be\label{betastar} \beta^\ast
= \frac{4\pi r_+}{\tilde d} \left(1 - \frac{r_-^{\tilde
d}}{r_+^{\tilde d}}\right)^{\frac{1}{\tilde d} - \frac{1}{2}}, \ee
which is the inverse of temperature at $\rho = \infty$.  The local
\be \label{localbeta} \beta = \Delta_+^{1/2}
\Delta_-^{-\frac{d}{2(D-2)}} \beta^{\ast}, \ee which is the inverse
of local temperature at $\rho$ when in thermal contact with
environment at the same temperature. For the canonical ensemble we
have fixed $\rho$ at the wall of cavity denoted by $\rho_B$, fixed
local temperature at $\rho_B$, fixed local brane volume $V_p =
\Delta_-^{\frac{\tilde d (d-1)}{2(D-2)}} V_p^\ast$, with $V_p^\ast =
\int d^p x$ and fixed charge defined as \be\label{charge} Q_d =
\frac{i}{\sqrt{2} \kappa} \int \ast F_{[p+2]} = \frac{\Omega_{\tilde
d +1}}{\sqrt{2} \kappa} \tilde d (r_+ r_-)^{\tilde d /2}. \ee In
\eqn{charge} $\kappa = \sqrt{8\pi G_D}$, where $G_D$ is the
$D$-dimensional Newton's constant, $\ast F_{[p+2]}$ denotes the
Hodge dual for the $(p+2)$ form-field given in \eqn{blackpbrane}.
Also $\Omega_n$ denotes the volume of a unit $n$-sphere. With these
data we will evaluate the action.

The relevant action for the gravity coupled to a $(p+1)$-form gauge
field in a manifold $M$ of dimension $D$ with the Euclidean
signature has the form, \be\label{action0} I_E = I_E(g) + I_E(F) \ee
where the first term is the purely gravitational action,
\be\label{action1} I_E(g) = - \frac{1}{2\kappa^2} \int_M d^D x
\sqrt{g} R + \frac{1}{\kappa^2} \int_{\partial M} d^{D-1} x
\sqrt{\gamma} (K-K_0) \ee consisting of the usual Einstein-Hilbert
term and the Gibbons-Hawking boundary term \cite{Gibbons:1976ue}
where $\partial M$ denotes the boundary of the manifold $M$. In the
above $K$ is the trace of the extrinsic curvature $K_{\mu\nu}$
defined as, \be\label{extcurv} K_{\mu\nu} =
-\frac{1}{2}\left(\nabla_\mu n_\nu + \nabla_\nu n_\mu\right), \quad
{\rm and\,\,so}, \quad K = -\nabla_\mu n^\mu \ee where $n^\mu$ is a
space-like vector normal to the boundary and is normalized as $n_\mu
n^\mu = 1$. Also $\gamma_{\a\b}$ is the boundary metric with
$\a,\,\b$ the indices of the boundary coordinates and $\gamma$ is
its determinant. $K_0$ is the subtraction term which serves as an
infra-red regulator so that a finite result can be obtained
\cite{Hawking:1982dh, York:1986it, Braden:1990hw,
Witten:1998qj}\footnote{In the charged case, though this flat space
is a solution of the corresponding equations of motion, it is not a
thermal state and serves only as an infra-red regulator. One way to
see the rationale behind is to first calculate the corresponding
action in the grand canonical ensemble for which the flat space is a
good reference state subtractor. Then the Euclidean action in the
canonical ensemble can be obtained from the calculated Euclidean
action in the grand canonical ensemble by a thermodynamical Legendre
transformation, noting that the respective action is related to
$\beta$ times the corresponding ensemble potential to leading order.
The canonical ensemble action so obtained  is also finite and is
actually equivalent to that by the subtraction procedure mentioned
in the text.}. This is calculated by embedding the same surface in
the flat space.

The second term is due to the form-field $F_{[p+2]}$ and its
expression in the canonical ensemble has the form,
\be\label{action2} I_E(F) = \frac{1}{2\kappa^2}
\frac{1}{2(d+1)!}\int_M d^D x \sqrt{g} F_{[d+1]}^2 -
\frac{1}{2\kappa^2} \frac{1}{d!}\int_{\partial M} d^{D-1} x
\sqrt{\gamma} n_\mu F^{\mu\mu_1\ldots \mu_d} A_{\mu_1\ldots \mu_d}
\ee Note that for canonical ensemble the charge in the cavity is
fixed.  For a given gravitational configuration in Euclidean
signature, the corresponding thermodynamical partition function in
the saddle point (or the zero-loop) approximation can be obtained as
$Z \approx e^{-I_E}$, where $I_E$ is evaluated for the given
configuration (see for example \cite{Brown:1994su}) and we will
evaluate it in the black $p$-brane configuration given in
\eqn{blackpbrane}. On the other hand, we also have $Z = e^{-\beta
F}$, where $\beta$ is the inverse temperature of the ensemble  and
$F$ is the corresponding free energy (for canonical ensemble this is
Helmholtz free energy). Therefore, we have in this approximation
$F=I_E/\b$. This is the relevant quantity we are going to evaluate
for studying the stability of the black branes and its phase
structure.

To evaluate $I_E$ given in \eqn{action0}, we can use the equation of
motion of the metric and obtain, \be\label{curvature} R =
\frac{\tilde d - d}{2(D-2)(d+1)!}F_{[d+1]}^2 \ee Substituting
\eqn{curvature} in the action we rewrite $I_E$ as,
\bea\label{actionred1} I_E &=& \frac{d}{2\kappa^2 (D-2)(d+1)!}\int_M
d^D x \sqrt{g} F_{[d+1]}^2 + \frac{1}{\kappa^2} \int_{\partial M}
d^{D-1} x \sqrt{\gamma}(K-K_0)\nn & & - \frac{1}{2\kappa^2}
\frac{1}{d!}\int_{\partial M} d^{D-1} x \sqrt{\gamma} n_\mu
F^{\mu\mu_1\ldots\mu_d} A_{\mu_1\ldots\mu_d} \eea In the above we
have used $d+\tilde d = D-2$ as mentioned earlier and $d\tilde d =
2(D-2)$ for the non-dilatonic branes which is a solution of
supergravity with maximal supersymmetry, the type we are
considering. We will evaluate each term in \eqn{actionred1}
separately. For that purpose we need the forms of the normal vector
$n^\mu$, the trace of the extrinsic curvatures $K$ and $K_0$ which
can be calculated from the metric in \eqn{blackpbrane} and are given
as, \bea\label{somequantity} n^\mu &=& \left(\Delta_+
\Delta_-\right)^{1/2} \delta^\mu_\rho\nn K &=& -\nabla_\mu n^\mu =
-\frac{1}{\sqrt{g}}\partial_\mu\left(\sqrt{g}n^\mu\right)\nn &=&
-\frac{\tilde d +1}{\rho}\left(\Delta_+ \Delta_-\right)^{1/2} -
\frac{\tilde d}{2 \rho^{\tilde d
    +1}}\left[\left(\frac{\Delta_-}{\Delta_+}\right)^{1/2} r_+^{\tilde d}
+  \left(\frac{\Delta_+}{\Delta_-}\right)^{1/2} r_-^{\tilde d}\right]\nn
K_0 &=& -\frac{\tilde d + 1}{\rho}
\eea
Now from the metric and the form-field given in \eqn{blackpbrane}, it is
straightforward to evaluate the action \eqn{actionred1} as,
\be\label{actionred2}
I_E = -\frac{\beta V_p \Omega_{\tilde d +1}}{2\kappa^2}\rho_B^{\tilde d}
\left.\left[2\left(\frac{\Delta_+}{\Delta_-}\right)^{1/2} + {\tilde d}
\left(\frac{\Delta_-}{\Delta_+}\right)^{1/2} + {\tilde d}\left(\Delta_+
  \Delta_-\right)^{1/2} - 2(\tilde d +1)\right]\right|_{\rho=\rho_B}
\ee Note in the above that $\rho$ is fixed on the cavity as
$\rho_B$. Now since we have $I_E = \beta F$, with $F=E-TS$, the
Helmholtz free energy in the canonical ensemble, so this implies
that $I_E = \beta E - S$, with $E$ the energy in the cavity enclosed
and $S$, the entropy for the system. Following  Braden et. al.
\cite{Braden:1990hw} for black hole, we expect that the entropy in
the present case (which is just that of the black brane under
consideration) should be independent of the choice of the location
of the cavity. However, this is not manifest in \eqn{actionred2}.
For this we can rewrite \eqn{actionred2} as, \bea\label{actionred3}
I_E &=& - \frac{\beta V_p \Omega_{\tilde d + 1}}{2 \kappa^2}
\rho_B^{\tilde d} \left.\left[(\tilde d + 2)
\left(\frac{\Delta_+}{\Delta_-}\right)^{1/2} + \tilde d (\Delta_+
\Delta_-)^{1/2} - 2 (\tilde d + 1)\right] \right|_{\rho=\rho_B}\nn
&\,& - \frac{4\pi \,V_p^* \Omega_{\tilde d + 1}}{2\kappa^2} \,
r_+^{\tilde d + 1} \left(1 - \frac{r_-^{\tilde d}}{r_+^{\tilde
d}}\right)^{ \frac{1}{2} + \frac{1}{\tilde d}}, \eea where we have
used the expression of $\beta^{\ast}$ given in \eqn{betastar}. The
last term in \eqn{actionred3} is precisely the entropy of the
non-dilatonic black $p$-brane \bea\label{entropy} S &=& \frac{4\pi
\,V_p^* \Omega_{\tilde d + 1}}{2\kappa^2} \, r_+^{\tilde d + 1}
\left(1 - \frac{r_-^{\tilde d}}{r_+^{\tilde d}}\right)^{ \frac{1}{2}
+ \frac{1}{\tilde d}}\nn &=& \frac{4\pi V_p \Omega_{\tilde d
+1}}{2\kappa^2} r_+^{\tilde d +1} \Delta_-^{-\frac{1}{2} -
\frac{1}{\tilde d}}\left(1 - \frac{r_-^{\tilde d}} {r_+^{\tilde
d}}\right)^{\frac{1}{2} + \frac{1}{\tilde d}}, \eea where we have
used $V_p = \Delta_-^{\frac{\tilde d(d-1)}{2(D-2)}} V_p^\ast$, and is
independent of the location of the cavity. We can also read off the
energy for the cavity as \be\label{energy} E (\rho_B) = - \frac{ V_p
\Omega_{\tilde d + 1}}{2 \kappa^2} \rho_B^{\tilde d}
\left.\left[(\tilde d + 2)
\left(\frac{\Delta_+}{\Delta_-}\right)^{1/2} + \tilde d (\Delta_+
\Delta_-)^{1/2} - 2 (\tilde d + 1)\right] \right|_{\rho=\rho_B}, \ee
which approaches the ADM mass at $\rho_B \to \infty$. In casting the
Euclidean action in the form of \eqn{actionred3}, we actually
treated the fixed $\beta$ at the wall of cavity, the inverse of
temperature of the environment, to be independent of $r_+$ for the
time being. Then the internal energy $E$ and the entropy $S$ are
functions of $r_+$ only when $\rho_B, V_p$ and $Q_d$ are all fixed
as in this ensemble.  The stability of the black $p$-branes can now
be discussed from the action \eqn{actionred3} when the second line
of (\ref{entropy}) for entropy is employed to which we turn in the
next section.

\section{Generalities  and stability of black $p$-branes}

We have mentioned that in the canonical ensemble the charge inside the cavity
given in \eqn{charge} is fixed. This implies that $r_-$ is not an independent
parameter and can be expressed in terms of $r_+$ as,
\be\label{rminus}
r_- = \left(\frac{\sqrt{2}\kappa Q_d}{\Omega_{\tilde d +1} \tilde
    d}\right)^{2/\tilde d}\frac{1}{r_+} = \frac{(Q_d^{\ast})^2}{r_+}
\ee
where we have defined $Q_d^{\ast} = [(\sqrt{2}\kappa Q_d)/(\Omega_{\tilde
  d+1}\tilde d)]^{1/\tilde d}$. With these, $\Delta_-$ will be given as,
\be\label{Deltaminus} \Delta_- = 1 -
\left(\frac{(Q_d^\ast)^2}{r_+\rho}\right)^{\tilde d} \ee We are now
considering the canonical ensemble for which the quantities $\beta$,
$V_p$, $\rho_B$ and $Q_d$ are all fixed (this implies that
$\beta^{\ast}$, $V_p^{\ast}$ are not fixed), and therefore the only
parameter which can vary is $r_+$. This implies that the entropy
will change with $r_+$, therefore if we vary the action
\eqn{actionred3} with respect to $r_+$ and set that to zero (that
is, at the stationary point), we must be able to recover $\beta$
(since the temperature is a conjugate variable to the entropy) and
this in turn will give a non-trivial check that the form of action
\eqn{actionred3} we have obtained is indeed correct. For this we
will replace $V_p^\ast$ by $V_p$ using the relation, $V_p =
\Delta_-^{\frac{1}{2}+\frac{d}{2(D-2)}} V_p^{\ast} =
\Delta_-^{\frac{1}{2}+\frac{1}{\tilde d}} V_p^{\ast}$ where we have
used $ d \,\tilde d = 2 (D - 2)$ in the second equality and so,
\eqn{actionred3} takes the form, \bea\label{actionred4} I_E &=& -
\frac{\beta V_p \Omega_{\tilde d + 1}}{2 \kappa^2} \rho_B^{\tilde d}
\left[(\tilde d + 2) \left(\frac{\Delta_+}{\Delta_-}\right)^{1/2} +
\tilde d (\Delta_+ \Delta_-)^{1/2} - 2 (\tilde d + 1)\right]\nn &\,&
- \frac{4\pi \,V_p \Delta_-^{-\frac{1}{2} - \frac{1}{\tilde d}}
\Omega_{\tilde d + 1}}{2\kappa^2} \, r_+^{\tilde d + 1} \left(1 -
\frac{r_-^{\tilde d}}{r_+^{\tilde d}}\right)^{\frac{1}{2} +
\vec{}\frac{1}{\tilde d}}. \eea In the expression of $\Delta_{\pm}$
here and below the variable $\rho$ must be replaced by constant
$\rho_B$ the size of the cavity. From $\partial I_E/\partial r_+ =
0$, we obtain after some simplification, \be\label{betadeter}
\left[\beta \tilde d - 4\pi r_+ \Delta_+^{1/2} \Delta_-^{-
\frac{1}{\tilde d}} \left(1 - \frac{r_-^{\tilde d}}{r_+^{\tilde
d}}\right)^{\frac{1}{\tilde d}- \frac{1}{2}}\right] \left[\tilde d +
2 + \left(\frac{\tilde d}{2} - \frac{\tilde d + 2}{2
\Delta_-}\right)\left(1 - \frac{r_-^{\tilde d}}{r_+^{\tilde
d}}\right)\right] = 0. \ee Since the second factor in the l.h.s. of
\eqn{betadeter} is greater than zero, we must have \be\label{beta}
\beta = \frac{4 \pi r_+}{\tilde d} \Delta_+^{1/2}
\Delta_-^{-\frac{1}{\tilde d}} \left(1 - \frac{r_-^{\tilde
d}}{r_+^{\tilde d}}\right)^{\frac{1}{\tilde d} - \frac{1}{2}} =
\Delta_+^{1/2} \Delta_-^{-\frac{d}{2(D - 2)}} \beta^*. \ee This is
precisely the correct form of $\beta$ given in (\ref{localbeta}),
which we obtain from the metric in \eqn{blackpbrane}. We will use
the relation \eqn{beta} to discuss the stability and the phase
structure of black $p$-branes. For this purpose let us rewrite
$\beta$ explicitly as, \be\label{betaf} \beta = \frac{4 \pi
r_+}{\tilde d} \left(1 - \frac{{Q_d^*}^{2\tilde d}}{r_+^{2 \tilde
d}}\right)^{\frac{1}{\tilde d}- \frac{1}{2}} \left(1 -
\frac{r_+^{\tilde d}}{\rho_B^{\tilde d}}\right)^{1/2} \left(1 -
\frac{{Q_d^*}^{2\tilde d}}{r_+^{\tilde d} \rho_B^{\tilde
d}}\right)^{- \frac{1}{\tilde d}}, \ee where we have used
\eqn{rminus}. Now following refs.
\cite{Braden:1990hw,Carlip:2003ne,Lundgren:2006kt}, we define \be
\label{parameters} x = \left(\frac{r_+}{\rho_B}\right)^{\tilde d}
\leq 1,\quad {\bar b} = \frac{\beta}{4\pi \rho_B}, \quad q =
\left(\frac{Q_d^*}{\rho_B}\right)^{\tilde d}, \ee where the
dimensionless parameters $\bar b$, $q$ are fixed\footnote{Since the
charge $Q_d$ as well as $Q^\ast_d$ denotes the corresponding
absolute value, so $q$ also denotes the absolute value. Therefore,
from now on, we use $q$ to denote the absolute value of the reduced
charge.} but $x$ is the only parameter which can change. Note that
the parameter $\bar b$ is related to the inverse of temperature of
the environment, $q$ is related to the charge and $x$ is related to
the horizon size. Also, as $(Q_d^{\ast})^2/r_+^2 = r_-/r_+ < 1$, $x
> q$. In terms of these parameters the above equation of state (or
thermal equilibrium condition) for $\beta$, \eqn{betaf}, can be
rewritten as, \be \label{bparameter} \bar b = b_q (x) \ee where
\be\label{bfunction} b_q (x) = \frac{1}{\tilde d} \frac{x^{1/\tilde
d} (1 - x)^{1/2}}{\left(1 -
\frac{q^2}{x^2}\right)^{\frac{1}{2}-\frac{1}{\tilde d}} \left(1 -
\frac{q^2}{x}\right)^{\frac{1}{\tilde d}}}. \ee In the above
\eqn{bfunction} we have used \be\label{redefinition} \Delta_+ = 1 -
x, \qquad \Delta_- = 1 - \frac{q^2}{x},\qquad 1 - \frac{r_-^{\tilde
d}}{r_+^{\tilde d}} = 1 - \frac{q^2}{x^2}. \ee If we define the
reduced Euclidean action as \bea\label{actionred5} \tilde I_E
&\equiv& \frac{2 \kappa^2 I_E}{4\pi \rho_B^{\tilde d + 1} V_p
\Omega_{\tilde d + 1}}\nn &=&- \bar b \left[(\tilde d + 2)
\left(\frac{1 - x}{1 - \frac{q^2}{x}}\right)^{1/2} + \tilde d (1 -
x)^{1/2}\left( 1 - \frac{q^2}{x}\right)^{1/2} - 2 (\tilde d +
1)\right]\nn &\,& - x^{1 + \frac{1}{\tilde d}} \left(\frac{1 -
\frac{q^2}{x^2}}{1 - \frac{q^2}{x}}\right)^{\frac{1}{2} +
\frac{1}{\tilde d}}, \eea one can show \be \label{phase}
\frac{\partial \tilde I_E}{\partial x} = f_q(x) \left[\bar b - b_q
(x)\right] \ee where $b_q (x)$ is as given above in \eqn{bfunction}
and \be\label{ffunction} f_q(x) = (1 - x)^{- 1/2} \left(1 -
\frac{q^2}{x}\right)^{- 1/2} \left[\tilde d + 2 - \frac{\tilde d +
2}{2} \left(\frac{1 - \frac{q^2}{x^2}}{1 - \frac{q^2}{x}}\right) +
\frac{\tilde d}{2} \left(1 - \frac{q^2}{x^2}\right)\right] > 0. \ee
The equation of state (\ref{bparameter}) and the reduced action
\eqn{actionred5} are the only relevant equations which we need for
the discussion of stability and phase structure for the black
branes. While we derive them for the non-dilatonic branes, they
remain true even for dilatonic branes as we will show in the
Appendix. In other words, the analysis given in the following
sections works  for both the non-dilatonic and dilatonic branes.
These two equations depend only on the parameters $\bar b, q, \tilde
d$ and the variable $x$. The only relevant dimensionality is $\tilde
d$ which remains unchanged under the so-called double-dimensional
reductions \cite{Duff:1987bx}. In other words, the black branes
related by this reduction will have the same stability and phase
structure at least in the approximation employed. For example, the
$D = 11$ M2 branes and the $D = 10$ fundamental strings share the
same above mentioned properties and so do  the other branes related
by this kind of reductions (see, for example, \cite{Duff:1994an}).

 Now from \eqn{phase} we find that (at the stationary point)
\bea\label{condition} \frac{\partial b_q (x)}{\partial x}
> 0 &\quad& \Rightarrow \qquad \frac{\partial^2 \tilde I_E}{\partial x^2} < 0,
\nn \frac{\partial b_q (x)}{\partial x} < 0 &\quad& \Rightarrow
\qquad \frac{\partial^2 \tilde I_E}{\partial x^2} > 0, \eea
Therefore, the system will be stable (at least locally) when
$\frac{\partial
  b_q}{\partial x} < 0$, corresponding to the minimum of the Euclidean action or
the free energy. On the other hand, if $\frac{\partial b_q}{\partial
x} > 0$, the corresponding Euclidean action is maximum and the free
energy is also maximum and the system will be unstable. Since the
phase structure for the black $p$-branes are quite different for the
chargeless case and charged case, we will discuss them separately in
the next two sections.

\section{Chargeless  case}

In this section we will discuss the stability and the phase
structure for the chargeless black $p$-brane. So, we will put $q=0$
in all the expressions we obtained in section 3. The expression for
the reduced action now takes the form, \be\label{chargelessaction}
\tilde I_E = - 2 (\tilde d + 1) \bar b \left[ \left(1 -
x\right)^{1/2} - 1 \right] - x^{1 + \frac{1}{\tilde d}}, \ee whence
we have \be \label{chargelessphase} \frac{\partial \tilde
I_E}{\partial x} = f_0 (x) \left[\bar b - b_0 (x)\right]. \ee Where
$b_0(x)$ and $f_0 (x)$ are given as, \bea \label{chargelessfunction}
b_0 (x)  &=& \frac{1}{\tilde d}\, x^{1/\tilde d} (1 - x)^{1/2},\nn
f_0(x) &=& ( \tilde d + 1)  (1 - x)^{- 1/2} > 0. \eea The equation
of state is \be\label{eos} \bar b = b_0(x). \ee At the stationary
point of the action, we have \be\label{stationary} \frac{\partial^2
\tilde I_E}{\partial x^2} = - f_0 (\bar x) \left.\frac{\partial b_0
(x)}{\partial x}\right|_{x = \bar x}, \ee where ${\bar x}$ is a
solution of the equation \eqn{eos}. Since $b_0(x)=0$ at $x=0,\,1$
and $b_0(x) > 0$ between $0 < x < 1$, so $b_0(x)$ has a maximum in
between which can be determined from $\partial b_0(x)/\partial x =
0$ and has the value, \be\label{bmax} b_{\rm max} =
\frac{1}{\sqrt{2\tilde d}} \left(\frac{2}{\tilde d +
2}\right)^{\frac{1}{2} + \frac{1}{\tilde d}} \quad \Rightarrow \quad
T_{\rm min} = \frac{\sqrt{2 \tilde d}}{4\pi \rho_B}
\left(\frac{\tilde d + 2}{2}\right)^{\frac{1}{2}+\frac{1}{\tilde d}}
\ee at \be\label{xmax}
 x_{\rm max} =
\frac{2}{\tilde d + 2} \qquad \Rightarrow \qquad r_{+\,{\rm max}} =
\left(\frac{2}{\tilde d +2}\right)^{1/\tilde d}\rho_B \ee So, there
exists a maximum $b_{\rm max}$ (or minimum temperature $T_{\rm
  min}$) above (or below) which the system can not be in a black brane phase.
Now since $\partial \tilde I_E/\partial x > 0$ (from
\eqn{chargelessphase}) and $I_E = 0$ at $x=0$ (from
\eqn{chargelessaction}), therefore the system favors the `hot flat
space'. Note that for the four dimensional black holes $D=4$, and
$\tilde d = d =1$, and so, $T_{\rm min}= 3\sqrt{3}/(8\pi \rho_B)$
and $r_{+\, {\rm max}} = (2/3) \rho_B$ match exactly with the values
found in \cite{York:1986it, Lundgren:2006kt}. We will have the same
$T_{\rm min}$ and $r_{+\, {\rm max}}$ for black strings in $D = 5$,
black membranes in $D = 6$, up to black D6 branes in $D = 10$ since
these branes are related to the four dimensional black hole via the
double-dimensional reductions.

\begin{figure}
 \psfrag{A} {$0$} \psfrag{B}{$x_1$} \psfrag{C}{$x_{\rm max}$}
 \psfrag{D}{$x_g$} \psfrag{E}{$x_2$}
 \psfrag{F}{$1$}\psfrag{G}{$x$} \psfrag{H}{$\bar b$}
 \psfrag{K}{$b_{\rm max}$} \psfrag{L}{$b_0 $}
\begin{center}
  \includegraphics{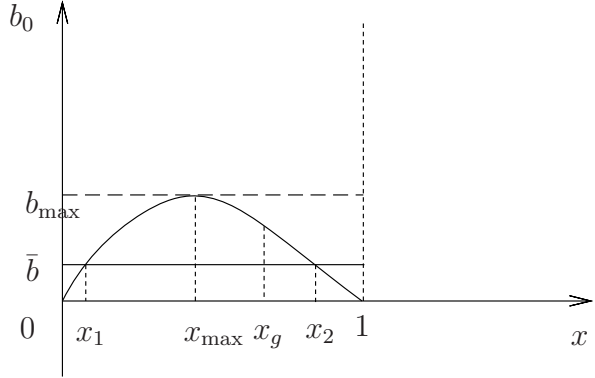}
  \end{center}
  \caption{The typical behavior of $b_0 (x) $ vs $x$ in the chargeless case.}
\end{figure}

For $\bar b$ smaller than $b_{\rm max}$, we can have two solutions
from the equation of state \eqn{eos}, but only the large solution
$x_2\,\,(>x_{\rm
  max})$ will be locally stable given the relation \eqn{stationary}.
This behavior of $b_0(x)$ versus $x$ is depicted in Figure 1.
However, this does not necessarily imply that the system is in the
black brane phase. Only when the local stability becomes a global
one, then the system is indeed in the black brane phase. The
corresponding $x_2$ can be determined from requiring the action at
the stationary point be negative. The action \eqn{chargelessaction}
can be expressed using \eqn{eos} as, \be\label{actionstationary}
\tilde I_E = - \frac{(\tilde d + 2)\bar b}{y} \left(y - \frac{\tilde
d}{\tilde d + 2}\right) (y - 1), \ee where we have defined
\be\label{yeqn} y = \sqrt{1 - \bar x} \ee with $x_2 = \bar x$. So,
the necessary condition for the global stability can be seen from
\eqn{actionstationary} to be \be\label{global}
 y < \frac{\tilde d}{\tilde d + 2}.
\ee This gives \be\label{xcondition} \bar x > x_g = \frac{4(\tilde d
+ 1)}{(\tilde d + 2)^2} > x_{\rm max}. \ee Now we find
\be\label{bcondition} b_g (x = x_g) = \frac{1}{\tilde d +
2}\left(\frac{4 (\tilde d + 1)}{(\tilde d + 2)^2}\right)^{1/\tilde
d} \quad \Rightarrow \quad T_g = \frac{(\tilde d +2)}{4\pi\rho_B}
\left(\frac{(\tilde d +2)^2}{4(\tilde d +1)}\right)^{1/\tilde d}.
\ee So only when $ 0 < \bar b < b_g$ (in this case $ x_g < x_2 <
1$), the system is in the black brane phase. On the other hand, for
$b_g < \bar b < b_{\rm max}$ (in this case $ x_{\rm max} < x_2 <
x_g$), the system, though locally stable, will eventually tunnel to
the `hot flat space' at the same temperature. However, at $\bar b =
b_g$, $\tilde I_E = 0$ and so, both the black brane phase with $\bar
x = x_g$ and the `hot flat space' phase are possible.  In other
words, this is the place at which the two phases can coexist and the
corresponding temperature is the phase transition one. This phase
transition is both a topological and a first order one since both
the topology
 and the entropy of the two phases are different before and after the phase
 transition. Note that when $x \to 1$,  $\bar b \to 0$ as
required from \eqn{eos}, but now $\bar b /y \to 1/\tilde d$, so the
action is still finite and is $\tilde I_E = - 1$, implying that the
system is stable. For the four dimensional black hole we find that
the temperature ($T_g$) where the large black brane becomes globally
stable is $27/(32 \pi \rho_B)$ which matches with the value found in
\cite{York:1986it, Lundgren:2006kt}. Also it is clear from Figure 1
that, as $b_0 (x)$ decreases or the temperature increases the size
of the small black brane decreases and that of the large black brane
increases and eventually when $b_0 \to 0$ or $T \to \infty$, the
size of the small black brane goes to zero and the size of the large
black brane approaches the size of the cavity. The phase transition
we found in the present case is analogous to the Hawking-Page
transition for the AdS black holes.

\section{Charged case}

In this section we will study the  stability and phase structure of
black $p$-brane in the more general case where the charge enclosed
by the cavity is non-zero and fixed. In section 3, while discussing
the generalities for the stability of black $p$-brane we found that
the system will be stable when $\partial b_q (x)/\partial x < 0$ and
will be unstable when $\partial b_q (x)/\partial x
> 0$. We will show that there exists a critical charge $q_c$, such that when
$q > q_c$, the system will be globally stable as $\partial b_q
(x)/\partial x$ is always less than zero (see Figure 3 below), but
when $q < q_c$, $\partial b_q (x)/\partial x > 0$ in some region as
shown in Figure 2. We mentioned in section 3 that the parameter $x$
lies between 1 and $ q$.  Since $b_q (x) \to \infty$ as $x \to q$
and $b_q (x) \to 0$ as $x \to 1$ (see eq.\eqn{bfunction}),
therefore, when $q < q_c $, $b_q (x) $ does not decrease
monotonically (as seen from Figure 2) and $\partial b_q (x)
/\partial x > 0$ in some region of $q < x < 1$. In other words, we
should have a minimum of $b_q (x)$ ($b_{\rm min}$) occurring at
$x=x_{\rm min}$ and a maximum of $b_q (x)$ ($b_{\rm max}$) occurring
at $x = x_{\rm max}$. When $b_{\rm min} < b_q (x) < b_{\rm max}$ and
$x_{\rm min} < x < x_{\rm max}$, $\partial b_q (x)/\partial x > 0$
and the system is unstable in this region and there exist no stable
brane  phases. On the other hand, when $b_q (x)
>b_{\rm max}$ or $b_q (x) <b_{\rm min}$, $\partial b_q (x)/\partial
x < 0$ and the system is stable (at least locally). So, for a given
$\bar b$ with $b_{\rm min}<\bar b <b_{\rm max}$, there will be three
solutions to the equation of state \eqn{bparameter}. If we denote
the three solutions as $x_1$, $x_2$ and $x_3$ with $x_1 < x_2 <
x_3$, then $x_1$ and $x_3$ correspond to the local minima of the
free energy and $x_2$ corresponds to the maximum. Among the two
minima, one expects that the one with the lower free energy will be
globally stable and a transition will occur from the state of higher
free energy to the lower free energy. So, it is important to find
which of the two black branes have lower free energy. To determine
this we write from \eqn{phase}, \be\label{freeenergy} \tilde I_E
(x_3) - \tilde I_E (x_1) = S(x_2, x_1) - \bar S (x_3, x_2), \ee
where \bea\label{sfunction} S (x_2, x_1) &=&
\int_{x_1}^{x_2} d x f_q (x) \left[\bar b - b_q (x)\right] \geq 0,\\
\bar S(x_3, x_2) &=& \int_{x_2}^{x_3} d x f_q (x) \left[b_q (x) -
\bar b\right] \geq 0, \eea with $b_q (x)$ as given in \eqn{bfunction}.
Note that for $\bar b = b_{\rm
 min}$, the points $x_2$ and $x_1$ coincide and so $S(x_2,x_1) =
0$ and $\bar S(x_2,x_3)$ takes a maximum value. Similarly, for $\bar
b = b_{\rm max}$, the points $x_2$ and $x_3$ coincide and therefore,
$\bar S(x_2,x_3) = 0$ and $S(x_2,x_1)$ takes a maximum value. Thus
both the function $S(x_2,x_1)$ and $\bar S(x_3,x_2)$ change
continuously from 0 to their maximum value as $\bar b$ is varied
from $b_{\rm min}$ to $b_{\rm max}$. So, in between there must exist
a fixed value of $\bar b$ which we call $\bar b_t$ (inverse of which
is related to a phase transition temperature) for each given charge
$q < q_c$, for which $S(x_2,x_1) = \bar S(x_3,x_2)$ and therefore
the Euclidean action or the free energies of the two stable black
brane configurations of sizes $x_1$ and $x_3$ are the same. In other
words, this is a phase transition temperature where the two black
brane phases coexist and make a transition freely from one phase to
the other, much like a van der Waals-Maxwell liquid-gas phase
transition. This was also noticed earlier for asymptotically AdS, dS
and flat black holes in canonical ensemble
\cite{Chamblin:1999tk,Lundgren:2006kt,Carlip:2003ne}.

\begin{figure}
\psfrag{A}{$b_q $} \psfrag{B}{$b_{\rm max}$} \psfrag{C} {$\bar b$}
\psfrag{D}{$b_{\rm min}$} \psfrag{E}{$q$}\psfrag{F}{$x_{\rm
min}$}\psfrag{G}{$x_{\rm max}$} \psfrag{H}{$1$}\psfrag{L}{$x$}
\psfrag{a}{$x_1$}\psfrag{b}{$x_2$}\psfrag{c}{$x_3$}
\begin{center}
  \includegraphics{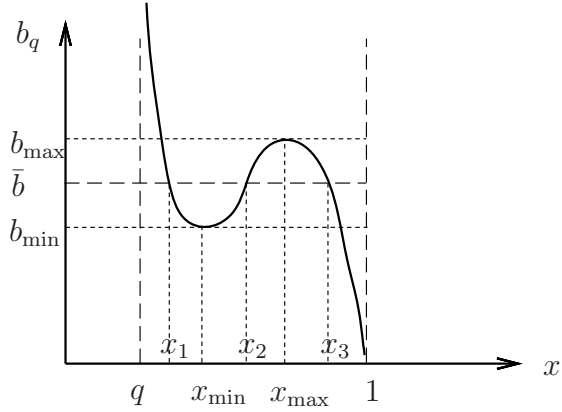}
  \end{center}
  \caption{The typical behavior of $b_q (x) $ vs $x$ when there is a phase
  transition ($q < q_c$).}
\end{figure}

 Now since at $b_{\rm max}$, $\bar S(x_3,x_2) =0$ and
$S(x_2,x_1)$ is maximum and at $\bar b_t$, they are the same, so, if
$\bar b$ lies in between i.e., $b_{\rm max}>\bar b > \bar b_t$,
$S(x_2,x_1) > \bar S(x_3,x_2)$, or $\tilde I_E(x_3) > \tilde
I_E(x_1)$. So, in this case the small black brane phase is globally
stable. Similarly, since at $b_{\rm min}$, $S(x_2,x_1) =0$ and $\bar
S(x_3,x_2)$ is maximum and at $\bar b_t$, they are the same, so, if
$\bar b$ lies in between i.e., $b_{\rm min}< \bar b < \bar b_t$,
$\bar S(x_3,x_2) > S(x_2,x_1)$, or $\tilde I_E(x_1) > \tilde
I_E(x_3)$. So, in this case large black brane phase is globally
stable. Thus we conclude that if the temperature is above the
transition value, but below $T_{\rm max}$, large black brane phase
is globally stable and if it is below the transition value, but
above $T_{\rm min}$, the small black brane phase is globally stable.
So, there will be a phase transition from the small black brane to
large black brane or {\it vice versa} depending on whether the
temperature of the black brane is below or above the transition
temperature.

It is clear from the expression of entropy given in the last term of
\eqn{actionred3} that for given $q < q_c$, the entropy will depend
on the parameter $r_+$ or $x$ and so, the entropy will change during
the phase transition we just mentioned. Therefore, this is a first
order phase transition for which the entropy has a discontinuity. In
fact there is a first order phase transition line when we move the
charge $q < q_c$ towards $q = q_c$ and the line gets shrunk ending
up at a second order phase transition point (critical point) for $q
= q_c$, which occurs at $x = x_c$, and where the entropy
discontinuity disappears.

Having understood the qualitative features of the equilibria and the
phase structure of black $p$-branes with non-zero charge inside the
cavity, we will try to understand the structure in a more
quantitative way and then corroborate our observations by numerical
calculations how the various situations for $q > q_c$, $q = q_c$ and
$q < q_c$ described here arise in three different values of $\tilde
d$, for examples, for the non-dilatonic branes (we will consider
only $\tilde d =3$ in $D = 11$, which corresponds to M5-brane,
$\tilde d =6$ in $D = 11$, which corresponds to M2-brane  and
$\tilde d = 4$ in $D = 10$, which corresponds to D3-brane) though
these calculations work for the corresponding dilatonic branes
related via the double-dimensional reductions in other dimensions as
well.

\begin{figure}
 \psfrag{A} {$b_q$} \psfrag{B}{$\bar b$}
 \psfrag{C}{$q$}\psfrag{D}{$\bar x$}\psfrag{E}{$1$}\psfrag{F}{$x$}
\begin{center}
  \includegraphics{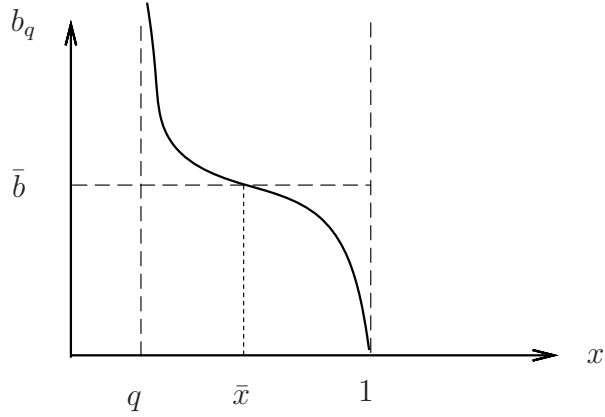}
  \end{center}
  \caption{ $b_q (x) $ decreases monotonically with $x$ when $q < x < 1$ ($q > q_c$).}
\end{figure}

 For understanding the stability, as we
mentioned, the quantity to look at is $\partial b_q (x) /\partial
x$. From \eqn{bfunction} we find, \be\label{delbdelx} \frac{\partial
b_q (x)}{\partial x} = - \frac{x^{1/\tilde d} \left\{(1 +
\frac{\tilde d}{2}) x^4 - [1 + (2 + \frac{\tilde d}{2})q^2 ]x^3 - 3
q^2 (\frac{\tilde d}{2} - 1) x^2 + q^2 [\tilde d - 1 + \frac{3
\tilde d}{2} q^2] x - \tilde d q^4\right\}}{{\tilde d}^2 \, x^4 \,(1
- x)^{1/2} \left(1 - \frac{q^2}{x^2}\right)^{1 + \frac{\tilde d}{2
(D - 2)}} \left(1 - \frac{q^2}{x}\right)^{1 + \frac{d}{2 (D - 2)}}}.
\ee The position of the extremality will be determined from the
vanishing of the numerator of the above equation \eqn{delbdelx},
i.e., \be\label{extremal} \left(1 + \frac{\tilde d}{2}\right) x^4 -
\left[1 + \left(2 + \frac{\tilde d}{2}\right)q^2 \right]x^3 - 3 q^2
\left(\frac{\tilde d}{2} - 1 \right) x^2 + q^2 \left[\tilde d - 1 +
\frac{3 \tilde d}{2} q^2\right] x - \tilde d q^4 = 0, \ee This is a
quartic equation and has four roots in general. We will make some
observation about the roots of this equation which will support the
various structures we described qualitatively in this section and
then give some numerical solution of this equation in some special
cases. First note that the discriminant of the above equation
\eqn{extremal} which tells us about the roots has the form,
\be\label{discriminant} \Delta (q, \tilde d) = - \frac{ (q^2 - 1)^3
q^6}{16}\left[ \big(4(\tilde d-1) - 3{\tilde d}(4+\tilde d)
q^2\big)^3 - 108\tilde d^2(2+\tilde d-\tilde d^2)^2
q^2(1-q^2)\right] \ee The discriminant will vanish within $0<q^2<1$,
if the last factor within the square bracket in \eqn{discriminant}
vanishes in that range. This will be determined by the intersection
point of the first term, i.e., a cubic curve with the second term,
i.e., a parabola. The parabola meets the $q^2$-axis at $q^2=0$ and
at $q^2=1$ and remains positive in this range. On the other hand the
cubic curve takes a positive value of $4^3 (\tilde d - 1)^3$ for
$\tilde d > 1$ at $q^2 = 0$, monotonically decreases and meets the
$q^2$-axis at $q_0^2 = 4(\tilde d -1)/(3\tilde d(4+\tilde d)) < 1$
(for $\tilde d
> 1$)\footnote{We will discuss $\tilde d =1$ case separately in a
subsection of this section.}.  Therefore, there is a unique crossing
point $q_c^2$ of the cubic curve and the parabola in the range
$0<q^2<1$ with $0< q^2_c < q_0^2 = 4(\tilde d -1)/(3\tilde
d(4+\tilde d)) < 1$. This shows the existence of a unique critical
point at $q = q_c$, where $\partial b_q (x) /\partial x$ vanishes.
Note that since this is a single extremum, as $b_q (x) $ varies from
$\infty$ to 0, this can not be a maximum or minimum, but is an
inflection point where $\partial^2 b_q (x) /\partial x^2$ also
vanishes. This feature is reflected in Figure 4 below.  Note that
for $\tilde d = 1$, $q_0^2 = 0$ and the intersection now occurs at
$q_0 = 0$ which is not in the range of  $0< q^2 < 1$, therefore we
don't expect a critical point and further a possible phase
transition to occur. This will be checked explicitly in a subsection
of this section later.

When $q^2>q_c^2$, we have $\Delta(q,\tilde d)<0$, so \eqn{extremal}
(since this is a quartic equation, there are four roots of this
equation in general), must have a pair of complex conjugate roots.
Also since the ratio of the coefficient of $x^4$ term and the
constant term (it is $- \tilde d q^4 /(1+\tilde d/2)$) is negative,
implying that the product of the four roots are negative, so one of
the roots must be negative. Hence, there can be at most one root in
the region $q < x <1$. However, for $\tilde d > 2$, as the l.h.s. of
equation \eqn{extremal} is positive at both $x=q$ (it is $(-2+\tilde
d)(-1+q^2)q^3$) and $x=1$ (it is $\tilde d(-1+q^2)^2/2$), the number
of roots, if existing at all in the region $ q < x <1$, must be
even\footnote{We will discuss $\tilde d =2$ in a separate subsection
of this section.}. So, there can not be any root of \eqn{extremal}
in the range $ q < x <1$ if $q^2 > q_c^2$ and this is consistent
with the typical behavior given in Figure 3.

When $q^2<q_c^2$, we have $\Delta(q,\tilde d)>0$ and this implies
that \eqn{extremal}  has either two pairs of complex conjugate roots
or all real roots. Also since the product of all four roots must be
negative, then the roots must all be real and the number of negative
roots must be odd (one or three). By the argument of previous
paragraph, the number of roots, if existing at all in the range $q <
x < 1$, must be even (two or none). The finding that $x_c$ falling
in the region $q < x <1$ for $q = q_c$, for each of the cases
considered in the following, implies that indeed there exist two
roots in the region $q < x <1$ for $q^2<q_c^2$. This is also
consistent with the typical behavior given in Figure 2.

Now we give some numerical calculation to illustrate the above
picture. When $\Delta(q,\tilde d) = 0$, one expects that $q^2$ in
\eqn{discriminant} has only one real positive root $q_c^2$, For this
critical $q_c$, we should have $x_{\rm min} = x_{\rm max}$ in the
region $q_c < x < 1$. Let us examine carefully if this is indeed
true. Let us first consider M5-brane, i.e., $\tilde d =3$ first. Now
$\Delta(q_c, \tilde d = 3) = 0$ gives us, \be\label{eq1} -512 +
27648 y - 110808 y^2 + 250047 y^3 = 0, \ee where we have defined $y
= q_c^2$. One can check that indeed $y$ has only one positive real
solution given by \bea\label{soln1} y &=& \frac{4 \left(114 - 4478
\left(\frac{2}{-450413 + 222607 \sqrt{5}} \right)^{1/3} + 2^{2/3}
\left(-450413 + 222607
   \sqrt{5}\right)^{1/3}\right)}{3087}\nn
   &\approx& 0.020058,
\eea whence we get \be\label{soln1a} q_c = 0.141626. \ee
Substituting this in \eqn{extremal} we obtain \be\label{eqn1a}
-0.00120697 + 0.0419264 x - 0.030087 x^2 - 1.0702 x^3 +
   \frac{5}{2} x^4 = 0,
\ee which indeed gives a unique solution in the region $q_c < x < 1$
as, \be\label{soln1b} x_c = 0.292675 \ee (the other two irrelevant
solutions are $x=-0.187352$ and $x=0.030083$). So, we now have the
critical $b_c$ as, \be\label{soln1c} b_c =
   \frac{x_c^{1/3} (1 - x_c)^{1/2}}{3 (1 - \frac{q_c^2}{x_c^2})^{1/6}
   (1 - \frac{q_c^2}{x_c})^{1/3}} = 0.199253.
\ee

\begin{figure}
 \psfrag{A} {$b_q $} \psfrag{B}{$b_c$} \psfrag{C}{$q$}
 \psfrag{D}{$x_c$} \psfrag{E}{$1$} \psfrag{F}{$x$}
\begin{center}
  \includegraphics{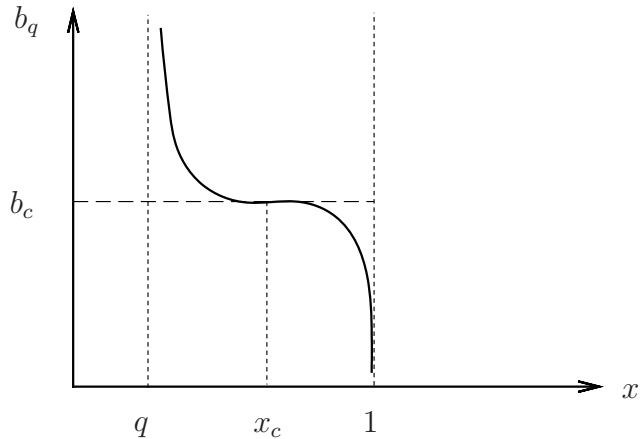}
  \end{center}
  \caption{The typical behavior of $b_q (x) $ vs $x$ when there is a turning point
for which $x_{\rm min} = x_{\rm max} = x_c$ ($q = q_c$).}
\end{figure}

From \eqn{extremal}, one expects that there exist a minimum of $b_q
(x)$, occurring at  $x = x_{\rm min}$ and a maximum, occurring at $x
= x_{\rm max}$, when $q < q_c $, corresponding to Figure 2
($\Delta(q,\tilde d) > 0$ in this case), while such $x_{\rm min}$
and $x_{\rm max}$ should not exist when $q
> q_c $ corresponding to Figure 3 ($\Delta(q, \tilde d) < 0$ in this
case). Let us take two explicit examples, one for each case, showing
that this is indeed true. When we take $q = 0.150000 > q_c$ which is
slightly larger than the critical value, we find indeed that there
exist no real solution in the region $q < x < 1$. While if we take
$q = 0.130000 < q_c$ which is slightly smaller than the critical
value, we find that indeed there exist two solutions in the region
$q < x < 1$, one is $x_{\rm min} = 0.229179$ and the other is
$x_{\rm max} = 0.341762$, as expected.

For M2-brane, $\tilde d =6$, we find exactly the same behavior. In
this case $\Delta(q_c,\tilde d =6) = 0$ gives (see
\eqn{discriminant}) \be\label{eqn2} -8000 + 3264192 y - 4992192 y^2
+ 5832000 y^3 = 0, \ee where again we have defined $y = q^2_c$. The
only real positive solution of \eqn{eqn2} gives, \bea y &=&
\frac{321 - 9506 \left(\frac{7}{-203167 + 168250
\sqrt{2}}\right)^{1/3} + 2 \, 7^{2/3} \left(-203167 + 168250
\sqrt{2}\right)^{1/3}}{1125},\nn
 &\approx& 0.00246007,\eea
which gives \be\label{soln2} q_c = 0.049599. \ee This in turn gives
the unique solution
   \be\label{soln2a}
x_c = 0.175176, \ee from the corresponding equation \eqn{extremal}
\be -0.0000363115 + 0.0123548 x - 0.0147604 x^2 - 1.0123 x^3 + 4 x^4
=0, \ee in the region of $q_c < x < 1$. We also have \be b_c =
0.116698. \ee The other cases $q > q_c$ and $q < q_c$ can be
discussed similarly as above.

For D3-brane, $\tilde d = 4$, $\Delta (q_c,\tilde d = 4) = 0$ gives
\be -1728 + 214272 y - 504576 y^2 + 884736 y^3 = 0, \ee whose unique
positive real solution is \bea y &=& \frac{ 73 - 1315
\left(\frac{5}{-34367 + 16512 \sqrt{6}}\right)^{1/3} +
   5^{2/3} \left(-34367 + 16512\sqrt{6}
   \right)^{1/3}}{384},\nn
   &\approx& 0.00822139,
   \eea
   which gives \be q_c = 0.090672. \ee With this critical $q_c $, we have
the unique \be x_c = 0.238800 \ee in the region of $q_c < x < 1$
from the corresponding equation \eqn{extremal} \be -0.000270365 +
0.0250697 x - 0.0246642 x^2 - 1.03289 x^3 + 3 x^4 = 0. \ee Now \be
b_c = 0.159921. \ee Once again, the other cases $ q > q_c$ and $q <
q_c $ can be similarly discussed.

We have given the numerical results for $\tilde d = 3,\, 6,\,$ and 4
since they are related to the M5, M2 and D3 branes. As indicated
earlier and with the discussion given in the Appendix for dilatonic
branes, any brane, non-dilatonic or dilatonic, related to each of
these branes by the so-called double-dimensional reductions will
share the same properties, since the only dimensionality entering
this discussion is  $\tilde d$ and it remains the same under this
reduction.

  Note that since we have $1 \le \tilde d \le 7$, for completeness and to show that
the critical value $x_c$ always falls in the range $q_c < x_c <1$,
we give the results also for $\tilde d = 5$ and 7 ($\tilde d =1,\,
2$ will be discussed separately). The critical values for $\tilde d
=5$ are $(q_c = 0.064944, \, x_c = 0.202012, \, b_c = 0.134632)$ and
for $\tilde d =7$ they are $(q_c = 0.039529,\, x_c = 0.154691,\, b_c
= 0.103210)$. We collect the relevant quantities for the cases $3
\le \tilde d \le 7$ in the tabular form as,

\vspace{.2cm}

\begin{center}
\begin{tabular}{|c|c|c|c|}
\hline\hline
$\tilde d$ & $q_c$ & $x_c$ & $b_c$\\
\hline\hline
3 & 0.141626 & 0.292656 & 0.199253\\
\hline
4 & 0.090672 & 0.238800 & 0.159921\\
\hline
5 & 0.064944 & 0.202012 & 0.134632\\
\hline
6 & 0.049599 & 0.175176 & 0.116698\\
\hline
7 & 0.039529 & 0.154691 & 0.103210\\
\hline\hline
\end{tabular}
\end{center}
\vspace{.2cm}

The table above shows that the critical size of the black brane
always lies in the range $q_c < x_c < 1$, where $q_c$ is related to
the absolute value of critical charge and $b_c$ is related to the
inverse critical temperature as defined earlier. We observe that the
critical quantities all decrease as $\tilde d$ increases.

\subsection{$\tilde d = 2$}

The general equilibria and the phase structure that we discussed in
this section does not apply to $\tilde d =2,\, 1$ cases as we
mentioned before and so, we will study these two cases separately
here in the next two subsections. Let us first discuss $\tilde d =
2$ case. We find from \eqn{bfunction} that for $\tilde d = 2$
\be\label{eq:bparameter-2} b_q (x) = \frac { x^{1/2} (1 -
x)^{1/2}}{2 \left(1 - \frac{q^2}{x}\right)^{1/2}}. \ee The
corresponding equation (from \eqn{extremal}) for finding the extrema
for $\tilde d =2$ is \be\label{extremal1} 2x^4 - \left(1 +
3q^2\right) x^3 + q^2 \left(1+3q^2\right)x - 2q^4 = 0 \ee which can
be factorized as, \be\label{factorform} (x^2-q^2)(x-x_+)(x-x_-) = 0
\ee where, \be\label{solution-2} x_{\pm} = \frac{1}{4}\left(1 + 3q^2
\pm \sqrt{\tilde \Delta}\right) \ee with $\tilde \Delta =
(1-q^2)(1-9q^2)$. The $x^2=q^2$ solutions are irrelevant here and
$x_{\pm}$ can be real only if $\tilde \Delta \geq 0$ implying
$0<q\leq 1/3$. Note that the discriminant for the present case from
\eqn{discriminant} has the form, \be\label{discriminant-2}
\Delta(q,2) = 4q^6(1-q^2)^3(1-9q^2)^3 \ee This gives the same
requirement as $\tilde \Delta$ for the reality of the solutions.
Comparing with our earlier discussion of $\Delta(q,\tilde d)$, we
note that here there is no parabola and the cubic curve meets $q^2$
axis at $q_0 = 1/3$. So, this is also the critical point $q_c = q_0
= 1/3$ which is consistent with the relevant solution of $\tilde
\Delta = 0$. Further we note that for this case, $b_q (x=q) =
\sqrt{q}/2$ which is different from $\infty$ (for non-zero charge),
the value for $\tilde d> 2$, but the other end $b(1) =0$ remains the
same.
\begin{figure}
 \psfrag{A} {$b_q $} \psfrag{B}{$\bar b$} \psfrag{C}{$q$}
 \psfrag{D}{$\bar x$} \psfrag{E}{$1$} \psfrag{F}{$x$}\psfrag{G}{$q >
 q_c$}
 \psfrag{a}{$b_c$}\psfrag{b}{$\bar b$}\psfrag{c}{$q$} \psfrag{d}{$\bar
 x$}\psfrag{e}{$1$}\psfrag{g}{$q = q_c$} \psfrag{Aa}{$b_{\rm
 max}$}\psfrag{Bb}{$\bar b$}\psfrag{Cc}{$q$}\psfrag{Dd}{$x_s$}
 \psfrag{Ee}{$x_+$}\psfrag{Ff}{$x_l$}
 \psfrag{Gg}{$1$}\psfrag{Hh}{$q < q_c$}
\begin{center}
\includegraphics{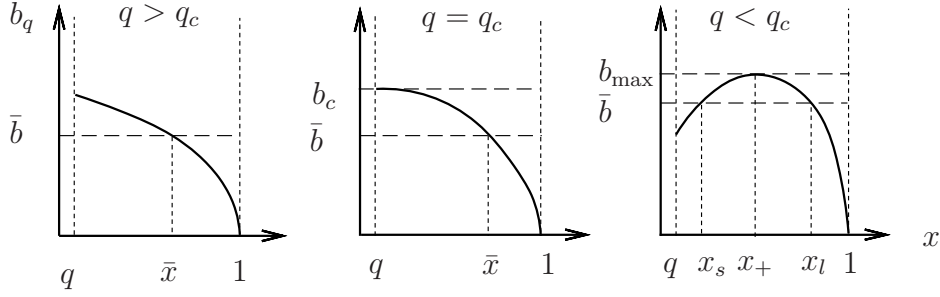}
  \end{center}
  \caption{The typical behaviors of $b_q (x)$ vs $x$ for $\tilde d = 2$. }
\end{figure}

Now let us see the phase structure in detail (see Figure 5). When $1
> q>q_c =1/3$, we see that there exists no extrema for $q<x<1$,
since now $\tilde \Delta < 0$ and \be\label{dbdxparameter-2} \frac
{\partial b_q (x)}{\partial x}=-\frac {b_q (x)} {2} \left[\frac
{2x^2-x(1+3q^2)+2q^2}{x(1-x)(x-q^2)}\right] < 0. \ee So, for now,
when $0< \bar b < b_q(q)$, we have a stable black
brane\footnote{Unless we have a phase like `hot flat space', but now
carrying a charge, whose free energy can be the lowest to be the
globally stable phase, this black brane phase is globally stable.}.
For $\bar b > b_q(q)$, there is no stable black brane and we don't
have a description available for such phase.

For $q=q_c=1/3$, $\tilde \Delta$ vanishes and the two roots
$x_{\pm}$ are equal and has the value $1/3 = q_c$ . So, $b_q (x=x_c)
= b_c$ is a fake critical point which is not accessible since this
is also an extremal point. This point is only marginally stable so
long as the thermodynamics is concerned. Now as we increase $x$
beyond this value, $b_q (x) $ monotonically decreases and goes to
zero at $x=1$ and in this range $\partial b_q /\partial x<0$. Thus
we find that in the range $b_c = 1/(2\sqrt{3}) > \bar b > b_q
(1)=0$, there exists a stable black brane phase. However, if $\bar
b> 1/(2\sqrt{3})$ there are no black brane phase or such description
is not available . In terms of temperature, if the temperature $T$
is below $\sqrt{3}/(2\pi\rho_B)$, there are no black brane phase,
however, if the temperature is in the range $\sqrt{3}/(2\pi\rho_B) <
T < \infty$, there is a stable black brane phase.

For $q< q_c = 1/3$, $\tilde \Delta > 0$, so in this case we have two
real solutions of $\partial b_q (x) /\partial x =0$ given by
\eqn{solution-2}. One can check that $x_-<q$ and $q< x_+ <1$ and so
only $x_+$ is the relevant solution which gives a maximum of $b_q
(x)$ at $x = x_+$ as  \be b_{\rm max}  = \frac{\left(1 + 3 q^2 +
 \sqrt{(1 - q^2)(1 - 9 q^2)}\right)^{1/2} \left(3\sqrt{1 - q^2} - \sqrt{1 - 9
 q^2}\right)}{8 \sqrt{2}} > b_q (q) = \frac{\sqrt{q}}{2}.\ee
Note that for $q < x < x_+$, $\partial b_q (x)/\partial x > 0$ while
for $x_+ < x < 1$, it is less than zero. Therefore in the range
$0<\bar b < b_q (q)=\sqrt{q}/2$, there is only one black brane phase
with $1
> x >x_+$ which is stable (we assume that there exists no other
stable phase) and in the range $b_q (q) < \bar b < b_{\rm max}$,
there are two black brane phases in which the smaller one is
unstable and the larger one is stable. For $\bar b > b_{\rm max}$
there is no black brane phase or such a description is not
available.

\subsection{$\tilde d =1$}

For $\tilde d =1$, we find from \eqn{bfunction} the form of the
parameter $b_q (x)$ as \be\label{eq:bparameter-1} b_q (x) =  \frac{x
(1 - x)^{1/2}\left(1 - \frac{q^2}{x^2}\right)^{\frac1 2 }}{ 1 -
\frac{q^2}{x}}. \ee From $b_q (x=q)= b_q (x=1)=0$ and the fact that
$b_q (x)
> 0$ for $q < x < 1$,  there must exist one and only one extremum
which corresponds to a maximum, denoted as $b_{\rm max}$ (see Figure
6). This is due to the absence of a critical point as discussed
earlier for the present case.  Also as discussed in the previous
section, since now $q_0^2 = 0$, so \be\label{dis-1} \Delta (q, 1) =
\frac{(15)^3 (q^2 - 1)^3 q^8}{16} \left[q^4 - \frac{16}{125} q^2 +
\frac{16}{125}\right] < 0, \ee for $0 < q^2 < 1$. So there must
exist a pair of complex conjugate solutions for the following
extremal equation of $b_q (x)$ which is a special case of
(\ref{extremal}) for $\tilde d = 1$, \be \label{extremalone}
\frac{3}{2} x^4 - \left(1 + \frac{5}{2} \,q^2\right) x^3 +
\frac{3}{2}\, q^2 x^2 + \frac{3}{2} \,q^4\, x - q^4 = 0.\ee
Furthermore, as the product of four roots of the above equation is
less than zero, this implies that the other two roots must be one
negative and one positive. Given that there exists a maximum, this
positive root must be the one we expect and should fall in the
region $q < x_{\rm max} < 1$. Let us take a special case of $q =
0.50$ as an example. Now the four solutions of (\ref{extremalone})
are \be x_1 = -0.30,\qquad x_2 = 0.33 - 0.29 i,\qquad x_3 = 0.33 +
0.29 i,\qquad x_4 = 0.73,\ee which are exactly as expected with only
$x_4 = 0.73 > q$ as the location of maximum of $b_q (x)$ for this $q
= 0.50$. The situation here is similar to the chargeless case but we
have here $q < x < 1$ for $b_q (x)$ vs $x$. In addition, we do not
have the corresponding transition to `hot flat space', i.e., the
Hawking-Page transition, since the charge is fixed for the present
system.

\begin{figure}
 \psfrag{A} {$b_q $} \psfrag{B}{$b_{\rm max}$} \psfrag{C}{$\bar b$}
 \psfrag{D}{$q$} \psfrag{E}{$x_s$} \psfrag{F}{$x_{\rm
 max}$}\psfrag{G}{$x_l$}\psfrag{H}{$1$} \psfrag{L}{$x$}
\begin{center}
  \includegraphics{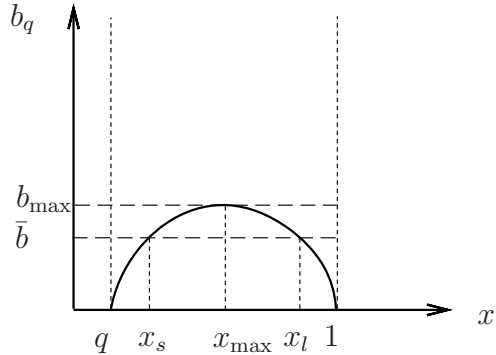}
  \end{center}
  \caption{The typical behavior of $b_q (x)$ vs $x$ for $\tilde d = 1$.}
\end{figure}

For a given $0<\bar b < b_{\rm max}$, there are two blackbrane
solutions: the small black brane is unstable while the large black
brane is at least locally stable. Unless some new phase carrying the
same charge is found with lower free energy, we can not justify
whether such a large brane is globally stable or locally stable.

This therefore completes our analysis of the equilibria and the
phase structure of black $p$-branes in the case of non-zero charge.
We have seen that the $\tilde d = 2$ serves as a borderline which
distinguishes the $\tilde d = 1$ case from the $\tilde d > 2$ cases.
For the former case, the introduction of charge doesn't change
qualitatively the stability behavior except for putting a new lower
bound on the $x$ set by the charge. In addition,  there doesn't
appear to exist the obvious analog of Hawking-Page type transition.
Apart from this, the phase structure remains basically the same as
the chargeless case. For the latter case, however, the introduction
of charge does significantly change the stability as well as the
phase structure from the chargeless case as described in detail in
this section. One of striking features is the appearance of a
critical charge which determines both the stability behavior and the
phase structure of the underlying system. When the charge is less
than the critical charge, there exists a first order phase
transition line which ends at a second order phase transition point
which is the critical point at the critical charge.  We would like
to remark that although the details of the phase structure is quite
different for $\tilde d > 2$ for black branes, the qualitative
structure is very similar to those of asymptotically AdS, dS and
flat black holes in canonical ensemble studied
earlier\cite{Chamblin:1999tk, Carlip:2003ne, Lundgren:2006kt}.

\section{Critical exponents}

We observed in section 5 for the case of charged black branes, that
when the charge is below certain critical value, $q < q_c$, there
exists two stable black brane states in certain range of
temperature. Let us denote the sizes of the two black branes as
$x_{s}$ and $x_{\ell}$, where the former denotes the small black
brane and the latter is the large black brane. There is a transition
temperature where the free energies of the two black branes are the
same and at this temperature these two black brane phases coexist.
This transition temperature, which is completely determined by the
given charge $q < q_c$, therefore, forming a first order phase
transition line, can be described by $T_t (q)$ (with the subscript
`t' denoting it as a phase transition temperature). As the charge
increases, this phase transition line ends up to a second order
phase transition point (critical point) at $q = q_c$. We would like
to calculate the critical exponents at this critical point.
Expanding $b_q (x)$ around the critical point $x_c$ we have,
\be\label{bexpansion} b_q - b_c = \frac{1}{3!}
\left.\frac{\partial^3 b_q}{\partial x^3}\right|_{x = x_c} (x -
x_c)^3 + \cdots. \ee Note that at the critical point the first and
the second order derivatives  of $b_q (x)$ with respect to $x$ are
zero. From \eqn{bfunction} we find, \be\label{thirdd}
\left.\frac{\partial^3 b_q}{\partial x^3}\right|_{x = x_c} = -
\frac{6 \,x_c^{\frac{1}{\tilde d}} \left\{2 \left(1 + \frac{\tilde
d}{2} \right) x_c^2 - \left[1  + \left(2 + \frac{\tilde d}{2}\right)
q_c^2\right] - q_c^2 \left(\frac{\tilde d}{2} -
1\right)\right\}}{{\tilde d}^2\, x^4_c \left(1 -
x_c\right)^{\frac{1}{2}} \left(1 -
\frac{q_c^2}{x_c^2}\right)^{\frac{\tilde d - 2}{2\tilde d} + 1}
\left(1 - \frac{q_c^2}{x_c}\right)^{\frac{1}{\tilde d} + 1}}, \ee
where we have used \be\label{firstsecond}
 \left.\frac{\partial b_q}{\partial x}\right|_{x = x_c} = 0,
\qquad \left. \frac{\partial^2 b_q}{\partial x^2}\right|_{x = x_c} =
0. \ee We mentioned in section 2, that the form of the entropy can
be read off from the last term of the Euclidean action
\eqn{actionred3}. If we now define the reduced entropy for fixed
$\rho_B$ and $V_p$ as, $\tilde S = 2\kappa^2 S/(\Omega_{\tilde d +1}
V_p \rho_B^{\tilde d +1})$, then from \eqn{actionred5} we read off
the form of reduced entropy as, \be\label{redentropy} \tilde S =
4\pi x^{\frac{\tilde d + 1}{\tilde d}} \left(1 -
  \frac{q^2}{x}\right)^{- \frac{1}{2} - \frac{1}{\tilde d}} \left(1
  - \frac{q^2}{x^2}\right)^{\frac{1}{2}+\frac{1}{\tilde d}}.
\ee We now expand the entropy around the critical point as,
\be\label{Sexpansion} \tilde S - \tilde S_c = \left.\frac{\partial
{\tilde S}}{\partial x}\right|_{x = x_c} (x - x_c) + \cdots, \ee
where from \eqn{redentropy}, \be\label{dsdx} \frac{\partial {\tilde
S}}{\partial x} = \frac{2\pi x^{\frac{1}{\tilde d}} \left[2 (\tilde
d + 1) x^3 - (3\tilde d + 4) q^2 x^2 + 2 q^2 x + q^4 \tilde
d\right]}{\tilde d x^3 \left(1 - \frac{q^2}{x}\right)^{\frac{3}{2} +
\frac{1}{\tilde d}} \left(1 - \frac{q^2}{x^2}\right)^{\frac{\tilde d
- 2}{2 \tilde d}}}. \ee We would like to have the expansion of
entropy not around the $x_c$, rather around the reduced critical
temperature $\tau_c=1/b_c$. Note that near the critical temperature,
\bea\label{tauexpansion} \tau-\tau_c &=& \frac{1}{b_q} -
\frac{1}{b_c}\nn &=& -\frac{1}{b_c^2}
\frac{1}{3!}\left.\frac{\partial^3 b_q}{\partial
    x^3}\right|_{x=x_c} (x-x_c)^3 + \cdots.
\eea where we have used \eqn{bexpansion}. So, using
\eqn{tauexpansion} we can write \eqn{Sexpansion} in terms of the
reduced temperature as, \be\label{newSexpansion} \tilde S - \tilde
S_c = \left.\frac{\partial {\tilde S}}{\partial x}\right|_{x = x_c}
\left[\frac{b_c^2}{- \frac{1}{3!} \left.\frac{\partial^3
b_q}{\partial x^3}\right|_{x = x_c}}\right]^{1/3} (\tau -
\tau_c)^{1/3} + \cdots. \ee The reduced specific heat therefore can
be calculated as, \bea\label{spheat} \tilde c_v &=& T \frac{\partial
\tilde S}{\partial T} = \tau \frac{\partial \tilde S}{\partial
\tau}\nn &=& \frac{1}{3} \left.\frac{\partial {\tilde S}}{\partial
x}\right|_{x = x_c} \left[\frac{1}{- \frac{1}{3!}
\left.\frac{\partial^3 b_q}{\partial x^3}\right|_{x = x_c} b_c}
\right]^{1/3} (\tau - \tau_c)^{- 2/3} + \cdots. \eea Therefore, the
critical exponent $\alpha$ of $\tilde c_v$ is $-2/3$ for black
branes when $\tilde d \ge 2$.

Let us consider M5 as an example. We can expand (\ref{bparameter})
around the critical value as \bea b_q - b_c &=& \frac{1}{3!}
\left.\frac{\partial^3 b_q}{\partial x^3}\right|_{x = x_c} (x -
x_c)^3 + \cdots\nn &=& -1.88444\, (x - x_c)^3 + \cdots.\eea \be
\tilde S - \tilde S_c = 12.6268 (x - x_c) + \cdots.\ee We have now
\be \tau - \tau_c = 47.465 (x - x_c)^3 + \cdots,\ee \be \tilde S -
\tilde S_c = 3.48741 (\tau - \tau_c)^{1/3} + \cdots,\ee and \be
\tilde c_v = 5.83415 (\tau - \tau_c)^{- 2/3} + \cdots. \ee Note that
the critical exponent for the specific heat has a universal value
$-2/3$ as was also noted for the asymptotically AdS, dS and flat
black holes earlier in \cite{Chamblin:1999tk,Carlip:2003ne}.

As indicated above already, the present analysis works  for both the
 non-dilatonic  and the dilatonic branes.

\section{Conclusion}

To conclude, in this paper we have studied in detail the equilibrium
states and the phase structures of the asymptotically flat
non-dilatonic and the dilatonic black branes in a cavity in
arbitrary space-time dimensions $D$. Although we mostly concentrated
on the non-dilatonic branes, the whole discussion applies also to
the dilatonic branes as we give the details of how the analysis can
be carried over in the Appendix. We considered only the canonical
ensemble and so the charge inside the cavity and the temperature at
the wall of the cavity were held fixed. We employed the Euclidean
action formalism to compute the thermodynamics and the phase
structure of the black branes. There is a marked difference in the
phase structure when the charge enclosed in the cavity is zero and
non-zero. When the charge is non-zero, there is also a qualitative
difference in the phase structure when the $\tilde d > 2$, $\tilde d
= 2$ and $\tilde d < 2$.  We discussed them each separately.

For zero charge we found an analog of Hawking-Page transition even
for these higher dimensional black objects in asymptotically flat
background. So, we found for the zero charge case that there exists
a minimum temperature given in eq.\eqn{bmax}, below which there is
no black brane phase and here the system will be in `hot flat space'
phase. But above this temperature there exists two black brane phase
with different radii. The smaller one is unstable and the larger one
is locally stable. When the temperature lies between the minimum
value \eqn{bmax} and the value given in \eqn{bcondition}, the
locally stable black brane will eventually tunnel into `hot flat
space' since the latter configuration in this region has lower free
energy. There is a phase transition temperature given by
\eqn{bcondition} at which the black brane of size  $x_g$ given in
\eqn{xcondition} and the `hot flat space' can coexist. But above
this temperature, the larger black brane becomes globally stable and
the system can remain in this black brane phase. However, the
smaller black brane still remains unstable. As we increase the
temperature the size of the smaller black brane becomes smaller and
that of the larger black brane becomes larger. As the temperature
tends to infinity the size of the smaller black brane tends to zero
and the larger black brane approaches the size of the cavity.

When the charge enclosed in the cavity is fixed but non-zero, we
found that there exists a critical charge $q_c$ at or above which
there is a single globally stable black brane phase in the absence
of an analog of `hot flat space' or some other unknown
configurations with favorable free energy for the charged system.
When the charge is below this critical value, then there exists a
certain range of temperature denoted by $T_{\rm min}$ and $T_{\rm
max}$ below and above which there is a single globally stable black
brane phase. But within this range there are three black brane phase
with different radii. The largest and the smallest of which are
locally stable as they correspond to the local minima of the free
energy, but the intermediate one is unstable as it corresponds to
the maximum of free energy. We found that the values of the free
energies of the black brane phases depend on the temperature. In
fact there exists a unique transition temperature at which the free
energies of the largest and the smallest black brane become equal
and so these two phases can coexist at this temperature and can make
a phase transition freely from one phase to the other much like the
van der Waals-Maxwell liquid gas phase transition. Above this
transition temperature the larger black brane is globally stable and
below this temperature the smaller black brane is globally stable.
So, there is a first order phase transition from smaller black brane
to larger black brane or {\it vice versa} above or below the
transition temperature as the entropy of the system changes in this
transition. In fact when the charge increases from $q < q_c$ towards
$q = q_c$, there is a first order phase transition line which
eventually ends up in a second order phase transition point
(critical point) at $q = q_c$. At this critical point we have
calculated the critical exponents and found that the critical
exponent of specific heat has a universal value $-2/3$. We have
elaborated this phase structure both analytically and numerically to
illustrate the various situations. We found that this general phase
structure is valid only for $\tilde d > 2$, where $\tilde d $ is
related to the dimensionality of the black $p$-brane by $\tilde d
=D- p -3$. $\tilde d=2,\,1$ case has been considered separately in
subsections 5.1 and 5.2. $\tilde d =1$ case is very similar in
structure to the zero charge case except here $b_q$ vs $x$ curve
starts from $x=q$ instead of $x=0$ as in zero charge case. Also here
there is no analog of `hot flat space' since the system has non-zero
charge. For $\tilde d =2$, we found that there exists a critical
charge $q_c$ above which there is a single globally stable black
brane phase when the temperature of the system is above a certain
value, but below this value of the temperature there is no black
brane phase and we do not have a suitable description for this
phase. When the charge becomes the critical value $q_c$, again there
exists a temperature $T = \sqrt{3}/(2\pi \bar\rho_B)$ above which
there is a single globally stable black brane phase and below this
temperature there is no description available for the charged
system. When the charge is below $q_c$, there is a certain range of
temperature where there are two black brane phases, the smaller one
is unstable and the larger one is globally stable. Above this range,
there is a single globally stable black brane phase and below the
minimum temperature we do not have a description available. This
whole analysis works for both the non-dilatonic and dilatonic
branes.

The only dimensionality which is relevant to the stability and phase
structure is $\tilde d$ and this implies that the branes related via
the so-called double-dimensional reductions have the same stability
and phase structure at least in the leading order approximation
adopted. For example, this implies that the $D = 11$ M2 brane, the
$D = 10$ fundamental string and the $D = 9$ 0-brane all have the
same stability and phase structure, so do the $D = 11$ M5, the $D =
10$ D4, the $D = 9$ 3-brane, upto the $D = 6$ 0-brane, and so on. We
also observe that the critical quantities ($q_c, x_c, b_c$) all
decrease when $\tilde d$ increases from $2$ to $7$.

\section*{Acknowledgements:}

JXL would like to thank the participants of the advanced workshop
``Dark Energy and Fundamental Theory", supported by the Special Fund
for Theoretical Physics from the NSF of China with grant no:
10947203, for stimulating discussions. He acknowledges support by
grants from the Chinese Academy of Sciences, a grant from 973
Program with grant No: 2007CB815401 and a grant from the NSF of
China with Grant No : 10975129.

\section*{Appendix}

Here we will consider the dilatonic black $p$-brane solutions in $D$
space-time dimensions. For studying thermodynamics we give their form in
the Euclidean signature as,
\bea
 \label{blackbrane-phi}
d s^2 &=& \Delta_+\Delta_-^{- \frac{d}{D - 2}} d t^2 +
    \Delta_-^{\frac{\tilde d}{D - 2}} \sum_{i=1}^{d-1}(d x^i)^2 +
\Delta_+^{-1}
    \Delta_-^{\frac{a^2}{2\tilde d} - 1} d\rho^2 + \rho^2
\Delta_-^{\frac{a^2}{2\tilde d}} \Omega_{\tilde d +
    1}^2,\nn
    A_{[p+1]} &=& - i e^{a\phi_0/2}
\left[\left(\frac{r_-}{r_+}\right)^{\tilde
    d/2} - \left(\frac{r_- r_+}{\rho^2}\right)^{\tilde
    d/2}\right] dt \wedge dx^1\wedge \ldots \wedge dx^p,\nn
    F_{[p+2]} &\equiv&  dA_{[p+1]}
    = - i e^{a\phi_0/2} \tilde d \,
\frac{(r_- r_+)^{\tilde d / 2}}{\rho^{\tilde d  +
    1}} d\rho \wedge dt \wedge dx^1 \wedge \ldots \wedge dx^p,
\nn
e^{2(\phi-\phi_0)}&=& \Delta_-^a,
\eea
where $\Delta_{\pm}$ are as defined before in section 2. $\phi$ is the dilaton
and $\phi_0$ is its asymptotic value and related to the string coupling as
$g_s = e^{\phi_0}$. $a$ is the dilaton coupling given by\footnote{Note that
  the form of $a^2$ is fixed by supersymmetry, in the sense that
these are solutions of supergravity with maximal supersymmetry.},
\be\label{dilatoncoupl} a^2 = 4 - \frac{2 d\tilde d}{D-2} \ee In the
metric given in \eqn{blackbrane-phi}, the  Euclidean time is
periodic with periodicity $\beta^{\ast}$ given as,
\be\label{temp-phi} \beta^{\ast} = \frac{4\pi r_+}{\tilde d} \left(1
- \frac{r_-^{\tilde d}}{r_+^{\tilde d}}\right)^{\frac{1}{\tilde
d}-\frac 1 2} \ee This is the inverse temperature at $\rho =
\infty$. The local $\beta$ is given as, \be\label{localtemp-phi}
\beta = \Delta_+^{1/2}\Delta_-^{-\frac{d}{2(d+\tilde d)}}
\beta^{\ast} \ee which is the inverse of local temperature at
$\rho$. The black $p$-brane will be placed in a cavity with its wall
at  $\rho = \rho_B$. It is clear from the metric in
\eqn{blackbrane-phi} that the physical radius of the cavity is
\be\label{barrho} \bar \rho_B = \Delta_-^{\frac{a^2}{4 \tilde d}}
\rho_B, \ee while $\rho_B$ is merely the coordinate radius. It is
this $\bar \rho_B$ which we should fix in the following discussion
and not $\rho_B$ (as in the non-dilatonic case). Also we fix the
dilaton on the boundary which is the requirement of obtaining its
standard equation of motion from the action given later. In other
words, we fix the dilaton at $\bar \rho_B$, which indicates that the
asymptotic value of the dilaton is not fixed for the present
consideration and this is crucial for our discussion. By this
argument we also have \be\label{barr} \bar r_{\pm} =
\Delta_-^{\frac{a^2}{4\tilde d}} r_{\pm} \ee and $\bar r_{\pm}$ are
the proper parameters which we should use in the present context. In
terms of the `barred' variables $\Delta_{\pm}$ remain the same as
before, \be\label{bardelta} \Delta_{\pm} = 1 - \frac{r_{\pm}^{\tilde
d}}{\rho_B^{\tilde d}} = 1 - \frac{\bar r_{\pm}^{\tilde d}}{\bar
\rho_B^{\tilde d}} \ee For the canonical ensemble we have fixed
local temperature at the wall of the cavity, fixed local brane
volume $V_p = \Delta_-^{\frac{\tilde d(d-1)}{2(D-2)}} V_p^{\ast}$
and fixed charge defined as, \bea\label{charge-phi} Q_d &=&
\frac{i}{2\sqrt{\kappa}}\int e^{-a(d)\phi} \ast F_{[p+2]} =
\frac{\Omega_{\tilde d +1}}{2\sqrt{\kappa}}e^{-a\phi_0/2}\tilde d
(r_+ r_-)^{\tilde d/2}\nn &=& \frac{\Omega_{\tilde d +1} \tilde
d}{\sqrt{2}\kappa} e^{-a\bar \phi/2} (\bar r_+ \bar r_-)^{\tilde
d/2} \eea where in the last line we have expressed the asymptotic
value of the dilaton by the fixed dilaton $\bar \phi \equiv
\phi(\bar \rho_B)$ at the wall of the cavity from the relation given
in \eqn{blackbrane-phi} and then expressed $r_{\pm}$ by $\bar
r_{\pm}$ from \eqn{barr}.

With these data we will now evaluate the action.
The relevant action for the gravity coupled to the dilaton and a $(p+1)$-form
gauge field with the Euclidean signature has the form
\be\label{action-phi}
I_E = I_E(g) + I_E(\phi) + I_E(F)
\ee
where, $I_E(g)$ is the gravitational part of the action, $I_E(\phi)$ is the
action for the dilaton and $I_E(F)$ is the action for the form-field and are
given as,
\begin{eqnarray}\label{partaction-phi}
 I_E (g) &=& - \frac{1}{2 \kappa^2} \int_M d^D x \sqrt{ g }\, R +
   \frac{1}{\kappa^2} \int_{\partial M}  d^{D - 1} x \sqrt{\gamma}
   \,(K-K_0)\,,
\nonumber\\
I_E(\phi)&=&-\frac 1{2 \kappa^2} \int_M d^D x \sqrt{g}
\left ( -\frac 1 2 (\partial \phi)^2\right)\,,
\nonumber\\
I_E (F) &=& \frac{1}{2 \kappa^2} \frac{1}{2 (d + 1)!} \int_M
   d^D x \sqrt{g}\,e^{-a(d)\phi}\, F^2_{d + 1}
\nonumber\\
&&- \frac{1}{2 \kappa^2}
   \frac{1}{d!} \int_{\partial M} d^{D - 1} x \sqrt{\gamma}\, n_\mu
   \,e^{-a(d)\phi}\, F^{ \mu \mu_1 \mu_2 \cdots \mu_d} A_{\mu_1 \mu_2 \cdots
   \mu_d},
\end{eqnarray}
The various quantities in the above actions have already been defined in
section 2.
The equations of motion following from the action \eqn{action-phi} have the
forms,
\begin{eqnarray}\label{EOM}
R_{\mu\nu} - \frac{1}{2} g_{\mu\nu} R& =&\frac 1 2 \partial_\mu\phi\partial_\nu
\phi-\frac 1 4 (\partial\phi)^2\,g_{\mu\nu}
\nonumber \\
&& +\frac{1}{2} \frac{1}{d!}
 e^{-a(d)\phi}\left( F_{\mu \mu_1 \mu_2 \cdots mu_d}
F_\nu\,^{\mu_1 \mu_2 \cdots \mu_d}
      -\frac{1}{2 (d + 1)} g_{\mu\nu} F^2\right),
\\
\Box\phi&=& -\frac {a(d)}{2(d+1)!} e^{- a(d)\phi} F_{[d+1]}^2
\\
\nabla_{\mu_1}(e^{-a(d)\phi}F^{\mu_1\cdots \mu_{d+1}})&=& 0
\end{eqnarray}
Using the equation of motion, the action can be reduced to:
\begin{eqnarray}\label{actionred1-phi}
I_E &=& \frac{d}{2(D - 2) \kappa^2 (d + 1)!} \int_M d^D x
      \sqrt{g} e^{-a(d)\phi}F^2_{d + 1} + \frac{1}{\kappa^2} \int_{\partial M}
      d^{D - 1} x \sqrt{\gamma} (K - K_0) \nn
      &\,&- \frac{1}{2 \kappa^2}
   \frac{1}{d!} \int_{\partial M} d^{D - 1} x \sqrt{\gamma}\, n_\mu
   \,e^{-a(d)\phi}\, F^{ \mu \mu_1 \mu_2 \cdots \mu_d} A_{\mu_1 \mu_2 \cdots
   \mu_d}\, .
\end{eqnarray}
From the metric in \eqn{blackbrane-phi} we have
\be\label{normal}
n^\mu = \Delta_+^{1/2} \Delta_-^{\frac{1}{2}-\frac{a^2}{4\tilde d}}
\delta^\mu_\rho,
\ee
The extrinsic curvature for the $p$-brane can be calculated as before,
\bea\label{extrinsic-phi}
K &=& -\nabla_\mu n^\mu=-\frac 1 {\sqrt {g}}\partial_\mu \left ( {\sqrt
{g}}n^\mu\right)
\nonumber \\
&=&-\Delta_+^{\frac12}\Delta_-^{\frac12-\frac{a^2}{4\tilde
d}}\frac 1 \rho \left(1-\frac{a^2}{4} +\frac {\tilde d} 2 \Delta
_+^{-1}+\left(\frac {\tilde d } 2
+\frac{a^2}{4}\right)\Delta_-^{-1}\right)
\eea
The extrinsic curvature $K_0$ can be calculated as,
\be\label{extrinsic0-phi}
K_0 = -\frac{(\tilde d +1)\Delta_-^{-\frac{a^2}{4\tilde d}}}{\rho} =
-\frac{\tilde d +1}{\bar \rho}
\ee
where we have defined $\bar \rho = \Delta_-^{\frac{a^2}{4\tilde d}} \rho$ and
note that with this redefined $\rho$, $K_0$ takes exactly the same form as
in the non-dilatonic case.

Now calculating each term in the action \eqn{actionred1-phi}
separately as before we obtain, \be\label{actionred2-phi} I_E = -
\frac{\beta V_p \Omega_{\tilde d + 1}}{2 \kappa^2} \bar
\rho_B^{\tilde d} \left[ 2
\left(\frac{\Delta_+}{\Delta_-}\right)^{1/2} +\tilde
d\left(\frac{\Delta_-}{\Delta_+}\right)^{1/2}+ \tilde d (\Delta_+
\Delta_-)^{1/2} - 2 (\tilde d + 1)\right]. \ee Note that in the
above action everything is expressed in terms of the `barred'
parameters defined earlier instead of the original parameters.
Comparing the reduced action \eqn{actionred2-phi} with the
corresponding reduced action for the non-dilatonic branes
\eqn{actionred2} we find that they have exactly the same form in
terms of the redefined parameters. Once we have this form of the
action \eqn{actionred2-phi}, we can rewrite it as before in the form
$I_E = \beta E - S$ as, \bea\label{actionred4-phi} I_E &=& -
\frac{\beta V_p \Omega_{\tilde d + 1}}{2 \kappa^2} \bar
\rho_B^{\tilde d} \left.\left[(\tilde d + 2)
\left(\frac{\Delta_+}{\Delta_-}\right)^{1/2} + \tilde d (\Delta_+
\Delta_-)^{1/2} - 2 (\tilde d + 1)\right] \right|_{\bar \rho=\bar
\rho_B}\nn &\,& - \frac{4\pi \,V_p^* \Omega_{\tilde d +
1}}{2\kappa^2} \, r_+^{\tilde d + 1} \left(1 - \frac{r_-^{\tilde
d}}{r_+^{\tilde d}}\right)^{\frac{1}{2} +\frac{1}{\tilde d}}, \eea
where we have used \be\label{betastar-phi} \beta^* = \frac{4\pi
r_+}{\tilde d} \left(1 - \frac{r_-^{\tilde d}}{r_+^{\tilde
d}}\right)^{\frac{1}{\tilde d} - \frac{1}{2}}, \ee We can thus
identify the entropy \be\label{entropy-phi} S = \frac{4\pi
V_p^{\ast}\Omega_{\tilde d+1}}{2\kappa^2} r_+^{\tilde d +1}
\left(1-\frac{r_-^{\tilde d}}{r_+^{\tilde d}}\right)^{\frac{1}{2} +
  \frac{1}{\tilde d}}
\ee
and the energy of the cavity as,
\be\label{energy-phi}
E = - \frac{V_p \Omega_{\tilde d + 1}}{2 \kappa^2} \bar \rho_B^{\tilde d}
\left[(\tilde d + 2)
\left(\frac{\Delta_+}{\Delta_-}\right)^{1/2} + \tilde d
(\Delta_+ \Delta_-)^{1/2} - 2 (\tilde d + 1)\right]
\ee
Note that the entropy has exactly the same form as that of the non-dilatonic
brane and we find that the energy approaches the ADM mass at
$\bar \rho_B \to \infty$ as expected. We would like to remark that all the
quantities like energy, entropy and temperature in the cavity all have the
invariant forms in terms of `unbarred' (non-dilatonic case) and `barred'
(dilatonic case) coordinates. Energy expression given in \eqn{energy-phi} has
already the same form as can be compared with \eqn{energy}. Entropy given
in \eqn{entropy-phi} can be written as,
\bea\label{entropy-phia}
S &=& \frac{4\pi V_p^{\ast}\Omega_{\tilde d+1}}{2\kappa^2} r_+^{\tilde d +1}
\left(1-\frac{r_-^{\tilde d}}{r_+^{\tilde d}}\right)^{\frac{1}{2} +
  \frac{1}{\tilde d}}\nn
&=& \frac{4\pi V_p \Omega_{\tilde d +1}}{2\kappa^2} \bar r_+^{\tilde
d +1} \Delta_-^{-\frac{1}{2} - \frac{1}{\tilde d}}
\left(1-\frac{\bar r_-^{\tilde d}}{\bar r_+^{\tilde
d}}\right)^{\frac{1}{2} +
  \frac{1}{\tilde d}}
\eea Comparing \eqn{entropy-phia} with \eqn{entropy} we find that
indeed they have exactly the same form. Similarly we have from
\eqn{localtemp-phi} and \eqn{temp-phi} the expression for inverse
temperature as, \be\label{tempnewform} \beta = \frac{4\pi \bar
r_+}{\tilde d} \Delta_+^{\frac{1}{2}}\Delta_-^{-\frac{1}{\tilde d}}
\left(1-\frac{\bar r_-^{\tilde
      d}}{\bar r_+^{\tilde d}}\right)^{\frac{1}{\tilde d} - \frac{1}{2}}
\ee Comparing \eqn{tempnewform} with \eqn{beta} we again find that
they have exactly the same form. This therefore indicates that the
thermodynamical quantities of the non-dilatonic branes essentially
have the same structure as the dilatonic branes with the new
physical parameters. Now using the expression of charge
\eqn{charge-phi} we can write $\bar r_-$ in terms of $\bar r_+$ as,
\be\label{rpmrelation} \bar r_- = \left(\frac{\sqrt{2} \kappa
Q_d}{\Omega_{\tilde d +1} \tilde d} e^{a\bar
\phi/2}\right)^{\frac{2}{\tilde d}} \frac{1}{\bar r_+} = \frac
{(Q_d^{\ast})^2}{\bar r_+} \ee Where we have defined $Q_d^{\ast} =
[(\sqrt{2}\kappa Q_d e^{a\bar \phi/2})/ (\Omega_{\tilde d +1} \tilde
d)]^{1/\tilde d}$. Note that since the charge $Q_d$ is fixed inside
the cavity and the dilaton $\bar \phi$ is fixed at the wall of the
cavity $Q_d^{\ast}$ is also fixed. Therefore, $\bar r_-$ is not an
independent parameter, but is dependent on $\bar r_+$ as given in
\eqn{rpmrelation}. Now using \eqn{rpmrelation} we can write
\eqn{tempnewform} as, \be\label{betaf-phi} \beta = \frac{4 \pi \bar
r_+}{\tilde d} \left(1 - \frac{{Q_d^*}^{2\tilde d}}{\bar r_+^{\tilde
2 \tilde d}}\right)^{\frac{1}{\tilde d} - \frac{1}{2}} \left(1 -
\frac{\bar r_+^{\tilde d}}{\bar \rho_B^{\tilde d}}\right)^{1/2}
\left(1 - \frac{{Q_d^*}^{2\tilde d}}{\bar r_+^{\tilde d} \bar
\rho_B^{\tilde d}}\right)^{\frac{1}{\tilde d}}, \ee Now as before we
can define \be\label{parameters-phi} x = \left(\frac{\bar r_+}{\bar
\rho_B}\right)^{\tilde d} \leq 1,\qquad \bar b =
\frac{\beta}{4\pi\bar \rho_B}, \qquad q =
\left(\frac{Q_d^{\ast}}{\bar \rho_B} \right)^{\tilde d} \ee In terms
of these parameters \eqn{betaf-phi} takes the form,
\be\label{bparameter-phi} \bar b = \frac{1}{\tilde d} x^{1/\tilde d}
(1-x)^{1/2} \left(1-\frac{q^2}{x^2}\right)^{\frac{1}{\tilde d} -
\frac{1}{2}} \left(1 - \frac{q^2}{x}\right)^{-\frac{1}{\tilde d}}
\equiv b_q(x) \ee Comparing with \eqn{bparameter} and
\eqn{bfunction} we find that this equation \eqn{bparameter-phi} has
exactly the same form as for the non-dilatonic branes. The
expression of $b_q (x)$ was crucial for our analysis for the
equilibria and stability structure of the black $p$-branes. Since
the dilatonic branes have the same expression for $b_q (x)$ as the
non-dilatonic branes, the phase structure for the dilatonic branes
would be exactly the same as the non-dilatonic branes. The reduced
Euclidean action $\tilde I_E$ can be seen from \eqn{actionred4-phi}
to take exactly the same form as the non-dilatonic branes given in
\eqn{actionred5}, \eqn{phase} and \eqn{ffunction} in terms of the
new parameters \eqn{parameters-phi}.

So far, we have not addressed the issue regarding the validity of
using the effective action in describing the phase structure of
black $p$-branes throughout the parameter space considered\footnote{We
thank the anonymous referee for raising this concern.}. Here we will
address this for both non-dilatonic and dilatonic branes together.
For non-dilatonic branes, we need to keep the curvature of black
brane spacetime uniformly weak throughout the parameter space. For
dilatonic branes, in addition, we also need to keep the effective string
coupling uniformly weak. The effective string coupling can be
read from \eqn{blackbrane-phi} and the curvature can be calculated
from the metric given in the same equation as \bea \label{cc} e^\phi
&=& g_s \Delta_-^{a / 2},\nn R &=& \frac{\tilde
d^2}{2}\left[\frac{a^2}{4} \frac{\Delta_+}{\Delta_-}
\left(\frac{\bar r_-}{\bar \rho}\right)^{2 \tilde d} \frac{\bar r_+
\bar r_-}{\bar\rho^2} + \frac{d - \tilde d}{D - 2} \left(\frac{\bar
r_+ \bar r_-}{\bar \rho^2}\right)^{\tilde d + 1}\right]
\frac{1}{\bar r_- \bar r_+}, \eea where $g_s = e^{\phi_0}$ is the
asymptotic string coupling. Note that for the scalar curvature
each term in the square bracket is less than unity since $\bar \rho
\geq \bar r_+ > \bar r_-$, $\Delta_+ /\Delta_- < 1$, $a^2 /4 <
1$ and $(d - \tilde d)/(D - 2) < 1$. In other words, the square
bracket contributes at most a factor of order unity to the
curvature. So in order to keep the curvature uniformly weak, we need
to have \be l^2 R \sim \frac{l^2}{\bar r_- \bar r_+} = l^2
\left[\frac{\Omega_{\tilde d + 1}}{\sqrt{2} \kappa Q_d e^{a
\phi_0/2} \Delta_-^{a^2/4}}\right]^{2/\tilde d} \ll 1,\ee where
$l$ is the relevant length scale under consideration, for example,
it is the Planck scale $l_p$ in eleven dimensions or the string scale
$l_s$ in ten dimensions. Note that the charge quantization gives
$\sqrt{2} \kappa Q_d e^{a \phi_0/2}/\Omega_{\tilde d + 1} \sim N
l^{\tilde d}$ with the integer $N$ labeling the number of branes. So
the uniformly weak curvature condition is \be \label{weakc} N
\Delta_-^{a^2/4} \gg 1.\ee For M branes, $a = 0$ and a weak
curvature is the only requirement which can be satisfied when $N \gg
1$. For D3 branes, we need in addition a small $g_s$. For those
branes, it doesn't appear that there is any constraint on the
parameter space we considered. For dilatonic branes, the condition
\eqn{weakc} for weak curvature can easily be satisfied for
non-extremal branes, i.e., $r_+ > r_-$, for given large enough $N$
since $\Delta_-$ is finite for $\rho \ge r_+$. Now the effective
string coupling as given in \eqn{cc} can remain small for small
$g_s$ when $a > 0$ and can also remain small if $g_s$ is chosen to
be small enough for $a < 0$. For this case, the parameter $x$ used
in the text falls in the range $q < x \le 1$. If we consider
extremal branes, i.e., $r_+ = r_-$, we then have to limit to the
range $\rho > r_+ = r_-$ so that the curvature remains small.
This can also give a small effective string coupling even when $a <
0$ by the same token. Now we have the range of parameter space as $q
\le x < 1$. In other words, for dilatonic branes, we can consider
the edge state only at one end, i.e., either at $x = q$ or at $x =
1$, but not at both.

   The discussion presented in this paper corresponds to the single-scalar $\Delta
   = 4$ black branes given in \cite{Duff:1996hp} (do not confuse this $\Delta$ with the
   discriminant notation used in the main text of this paper).
   However,  we have investigated the other $\Delta = 3, 2 , 1$ single scalar
   black branes discussed in \cite{Duff:1996hp} as well and found that the basic phase
   structure remains the same as that of $\Delta
   = 4$ black branes even though the minimal value of $\tilde d$ at which there exists a
  critical charge $q_c$ depends on the value of the respective
  $\Delta$.

\end{document}